\documentclass{FrogStyle}
\usepackage{placeins}
\usepackage{graphicx}
\usepackage[bookmarks=false]{hyperref}
\begin{document}

\long\def\symbolfootnote[#1]#2{\begingroup%
\def\thefootnote{\fnsymbol{footnote}}\footnote[#1]{#2}\endgroup} 

%==============================================================================
% title page for few authors

\begin{titlepage}

%  \cmsnote{2009/000}
  \date{January 18th, 2009}

  \title{\vspace{2cm}\textsc{Frog}: The Fast \& Realistic \textsc{OpenGL} Displayer\vspace{1cm}}

  \begin{Authlist}           
        Lo\"ic~Quertenmont\symbolfootnote[1]{Funded by the "Fonds de la Recherche Scientifique" (FNRS)}, Vincent~Roberfroid
  
       \Instfoot{ucl}{Center for Particle Physics and Phenomenology (CP3) \\ Universit\'{e} catholique de Louvain \\  Chemin du cyclotron 2, B-1348-Louvain-la-Neuve - Belgium}       
       \vspace{1.0cm}
  \end{Authlist}

  \begin{abstract}
\textsc{Frog}~\cite{REF_FROG} is a generic framework dedicated to visualisation of events in high energy experiment.  It is suitable to any particular physics experiment or detector design.
The code is light ($<3~\textrm{MB}$) and fast (browsing time $\sim20$ events per second for a large High Energy Physics experiment) and can run on various operating systems, as its object-oriented structure (C++) relies on the cross-platform \textsc{OpenGL}~\cite{REF_OPENGL} and \textsc{Glut}~\cite{REF_GLUT} libraries.  
Moreover, \textsc{Frog} does not require installation of third party libraries for the visualisation.
This document describes the features and principles of \textsc{Frog} version 1.106, its working scheme and numerous functionalities such as: 3D and 2D visualisations, graphical user interface, mouse interface, configuration files, production of pictures of various format\footnote{Please, cite the \textsc{Frog} project when publishing \textsc{Frog} based pictures}, integration of personal objects, etc. Finally, several examples of its current applications are presented for illustration.

\vspace{2.0cm}

\begin{figure}[ht!]
  \centering
  \includegraphics[width=0.29\linewidth]{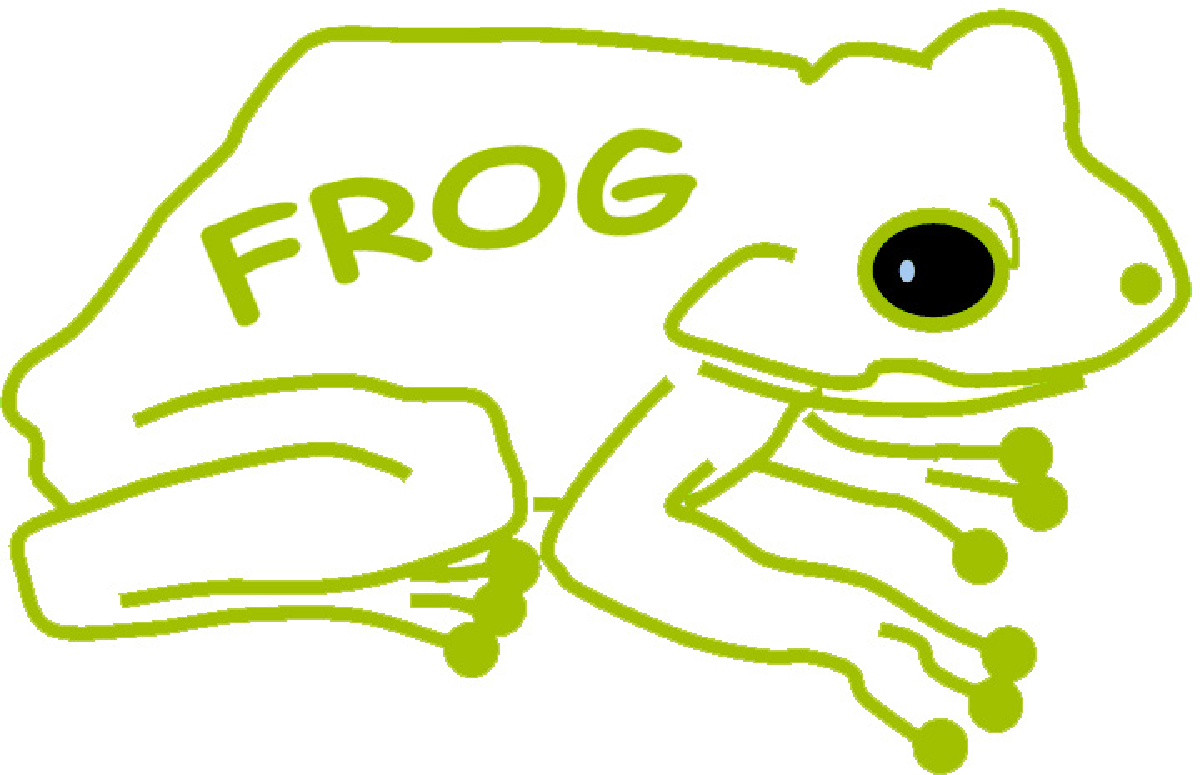}	  
\end{figure}

  \end{abstract} 
  
\end{titlepage}

\setcounter{page}{2}%JPP

\setcounter{tocdepth}{3}     % Dans la table des matieres
\setcounter{secnumdepth}{4}  % Avec un numero.
\setlength{\parindent}{0cm}

%==============================================================================

\tableofcontents
\newpage

%==============================================================================

\section{Introduction}

In high energy physics experiments, the possibility to visualise each event is crucial for several reasons. Understanding the event topologies can help in developing better data analysis algorithms.  
It can also be used as a powerful debugging tool for the simulation and reconstruction experiment softwares.

A visualisation tool has to be : 
\begin{itemize}
	\item fluid : draw $> 60~\textrm{frames~per~second}$
	\item fast : scan hundreds of events in few seconds
	\item light : the entire package should not exceed $10~\textrm{MB}$		
	\item easily upgradable	
	\item an intuitive debugging tool
	\item able to provide nice illustrations for communications
\end{itemize}

Satisfying simultaneously the above requirements could require heavy usage of all the resources of a computer. In general, these resources are the main limitation of the 3D visualisation.  In large experiments such as CMS~\cite{REF_CMS}, complex algorithms are used to reconstruct and analyse physics data. Since these algorithms are processor and memory consuming, a fast visualisation tool should be decoupled from simultaneous physics calculations.

\begin{figure}[hb!]
  \vspace{0.2cm}
	\centering
	\includegraphics[width=0.85\linewidth]{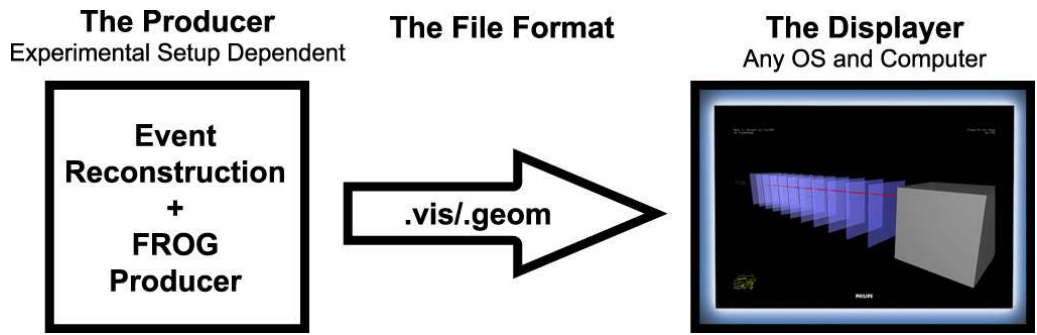}
	\caption{ Factorisation of \textsc{Frog} into the Producer and Displayer components.   The Producer is integrated in the experimental software.  It extracts, processes and stores the data required by the Displayer, using the \textsc{Frog} File Format for data encapsulation.   The Displayer is completely disconnected from the experimental software.}
  \label{fig:Introduction_Frog_Schematic}	
  \vspace{0.2cm}
\end{figure}

The \textsc{Frog}~\cite{REF_FROG} philosophy is to divide the software into a Producer and a Displayer (Figure~\ref{fig:Introduction_Frog_Schematic}).
It has an impact on several other elements of the software, like the input file format, the operating system portability and the structure of the software.
A dedicated \textsc{Frog} File Format (FFF) is needed in order to make the two parts of the code communicate with each other.
It has to be highly compact, but the decoding time of the files has also to remain as reduced as possible.
Finally, the FFF should also be flexible enough.
Its binary encoding ensures its universality with respect to the computer type and to the operating systems.  
The FFF is described in more detail in Section~\ref{sec:TheFileFormat}.

\

The Producer (Section~\ref{sec:Producer}) is the interface between the physics software of the detector/experiment and the \textsc{Frog} Displayer.  
This code depends on the respective experiment software but it does not require specific graphical libraries.
Its task is to extract and process once for all the data required by the Displayer.
For instance, it can convert the position of a hit from local coordinates of the detector to global coordinates.
Such computations are in general relatively fast but can slow down the 3D visualisation when repeated many times.
The Producer creates two separates files, one containing the processed geometry data (\texttt{.geom}) and one containing the needed events data (\texttt{.vis}).  
This is the part that the user has to adapt in order to make \textsc{Frog} working for his particular experiment.

The Displayer (Section~\ref{sec:Displayer}) has the unique function of displaying the content of the \texttt{.geom} and \texttt{.vis} files.
It is independent from the experiment software.
The Displayer can be distributed and run on different platforms : Windows, Linux and MacOS are currently supported.
The only requirement to execute the Displayer is to use some graphic libraries like \textsc{OpenGL}~\cite{REF_OPENGL} and \textsc{Glut}~\cite{REF_GLUT}.

\textsc{Frog} is completely generic in the sense that it has been designed to be usable for any experiment and also to allow everybody to easily upgrade/add some files in order to use \textsc{Frog} for his particular case.
It can then be used eventually for other purposes.
The two parts of the code, the Producer and the Displayer are described respectively in Section~\ref{sec:Producer} and~\ref{sec:Displayer}.
Additional features are described in Section~\ref{sec:features}.
A few applications of \textsc{Frog} in specific detector setups/environments are given in Section~\ref{sec:Applications}.
Future developments of the \textsc{Frog} Package are summarized in Section~\ref{sec:pespectives}. 
The two appendices~\ref{sec:base_classes}A and~\ref{sec:Complete_Example}B describe in detail the design and implementation of the software.

\section{\textsc{Frog} File Format}
\label{sec:TheFileFormat}

\begin{figure}[ht!]
  \begin{center}
    \includegraphics[width=0.4\linewidth]{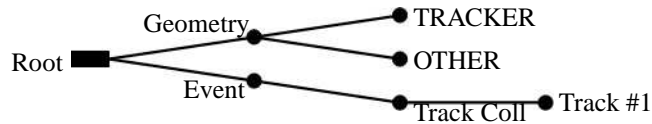}
    \put(-205, 15){Root}
    \put(-155, 30){Geometry}
    \put(-140,  5){Event}    
    \put(- 53, 32){TRACKER}    
    \put(- 53, 16){OTHER}    
    \put(- 53,- 4){Track Coll}    
    \put(   2,  0){Track \#1}        
    \caption{A schematic view of the tree structure of a FFF.}
    \label{fig:FROGCore_Chunk_Tree}
  \end{center}
\end{figure}

The \textsc{Frog} File Format (FFF) will first be described since it is of crucial importance and has a significant impact on the program structure.  
Since \textsc{Frog} has to be completely generic and has to store needed data in the most compact way, the dedicated file format is based on a binary encoding, where data are organised in \textit{chunks}.
Each chunk contains an \textit{Id} that specifies the chunk type, a \textit{Size} that indicates the chunk end, and the \textit{Data} themselves.  
The chunk \textit{Id} is written on $2~\textrm{Bytes}$, so $2^{16}=65~536$ different chunk types can be handled.
The \textit{Size} is written on $4~\textrm{Bytes}$, so the chunk size is limited to $2^{32}=4~\textrm{GBytes}$.  
The \textit{Size} is defined as : $$\rm{\textit{Size}} = \rm{sizeof(\textit{Id})} + \rm{sizeof(\textit{Size})} + \rm{sizeof(\textit{Data})} = 6 + \rm{sizeof(\textit{Data})}$$
The chunks can contain any type of data, which ensures a maximal flexibility to the software.
They can be divided into two categories : chunks that contains sub-chunks and those that don't.
By similarity with trees, the firsts are called "branches" while the others are called "leaves".  
The branches are useful to group together all data of a same experiment/detector/region.  See Fig.~\ref{fig:FROGCore_Chunk_Tree}.

\begin{figure}[ht!]
  \begin{center}
    \includegraphics[width=0.55\linewidth]{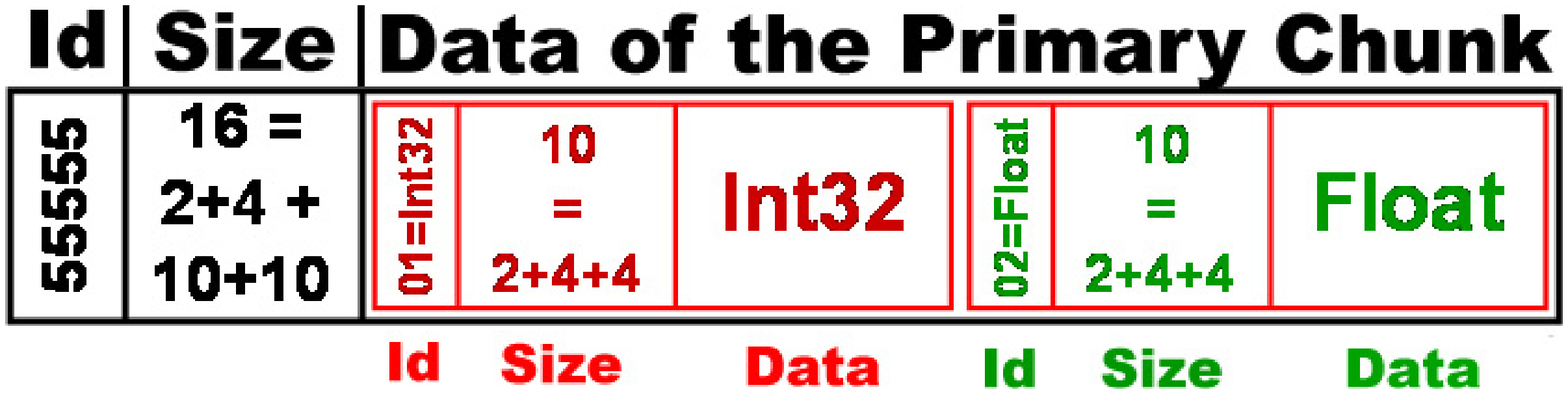}
    \caption{Example of data file. The primary chunk contains two sub-chunks storing an integer and a float.}
    \label{fig:FROGCore_Chunk_Example}
  \end{center}
\end{figure}

A file always contains a unique primary chunk that encapsulates, in its \textit{Data} part, all the other chunks, it's the root of the data tree.
It must be of type \texttt{C\_PRIMARY=55555}.  When the primary chunk \textit{Size} is not equal to the file size, the file is assumed to be corrupted.
The experiment data (e.g. the detector geometry or the event signals) are contained in the sub-chunks of this primary chunk.
The chunk structure is illustrated in Figure~\ref{fig:FROGCore_Chunk_Example}.\newline

%\begin{tabular}{|c|c|c|}
%\hline%\hline
%\bf{Id}    & \bf{Size}          & \bf{Data of primary chunk}   \\ \hline
%55555 &  16=2+4+10+10 &
%%%
%\begin{tabular}{cc}
%  \begin{tabular}{|c|c|c|}
%  %\hline
%  Id         & Size     & Data  \\ \hline
%  01 = Int32 & 10=2+4+4 & Int32 \\
%  %\hline
%  \end{tabular}
%&
%  \begin{tabular}{|c|c|c|}
%  %\hline 
%  Id         & Size     & Data  \\ \hline
%  02 = Float & 10=2+4+4 & Float \\
%  %\hline
%  \end{tabular}
%\end{tabular}
%%%%
% \\ \hline%\hline
%\end{tabular}

\begin{figure}[hb!]
  \begin{center}
    \includegraphics[width=0.55\linewidth]{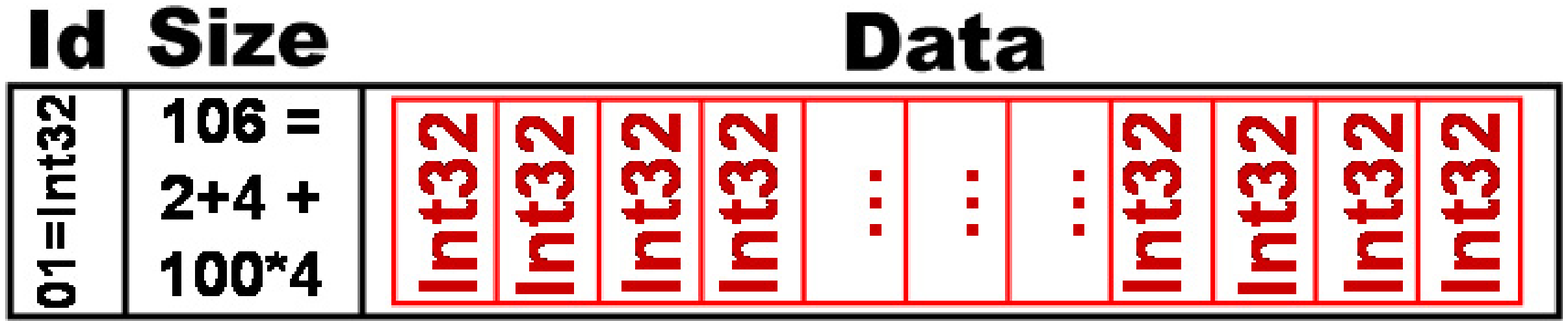}
    \caption{Example of the chunk structure if the file contains $100$ Int32.}
    \label{fig:FROGCore_Chunk_Example2}
  \end{center}
\end{figure}

It is often possible to optimize the way data are stored.
For instance, if a large number of data of the same type and the same size are stored (e.g. $100$ Int32) in a mother chunk. 
It is clear that it is only needed to define the size and the type once for all the data.
This optimized chunk is represented on Figure~\ref{fig:FROGCore_Chunk_Example2}.
The best storing method is automatically chosen by \textsc{Frog} in order to reduce the file size.
The definition of the chunk structures and the related methods can be found in the files: \texttt{FROG/FROG\_Chunk.cpp} and \texttt{FROG/FROG\_Chunk.h} files.\newline

\newpage
\section{\textsc{Frog} Producer}
\label{sec:Producer}

The \textsc{Frog} Producer builds the Events and the Geometry of a particular detector and stores them into \texttt{.vis} and \texttt{.geom} files.
It is the interface between the detector software\footnote{Two examples of detector software are CMSSW in the CMS experiment and Athena in the ATLAS experiment.} and \textsc{Frog}. A \textsc{Frog} Producer already exists for the CMSSW environment.
The Producer is the only part of \textsc{Frog} that has to be interfaced to the user needs, and to the experiment software and data format.

The Producer is in general a part of code (generally C++) that converts the experiment data format to the FFF.  
Since the Producer does not use specific graphical libraries it can be run in parallel on a computer cluster.  The many output \texttt{.vis} files can be merged in a unique file using the \textsc{Frog} merger included in the \textsc{Frog} package.  This tool is extremely fast since it just puts the event chunks spread into many files, in the same primary chunk of a unique file.

The experiment software only needs to include the textsc{Frog} classes definition in order to produce the \texttt{.vis} and \texttt{.geom}.  This reduced \textsc{Frog} package, containing only the class definition, takes of the order of $\sim~0.5\textrm{MB}$ on disk and is freely distributable.

A Producer example code of a fictitious experiment composed of a particle source beam, eleven tracking layers and a block to stop the beam, can be found in the appendix~\ref{sec:Complete_Example}.
This example is also available online as a tutorial~\cite{REF_FROGTUT}.

\section{\textsc{Frog} Displayer}
\label{sec:Displayer}

This part of the code is completely independent of the detector and does not need to be modified by the \textsc{Frog} users.
The code uses extensively the \textsc{OpenGL} library and is programmed such as to keep the rendering of events fast.

All the style parameters are loaded from the configuration file (\texttt{config.txt}). 
The geometry and the current event are displayed in the different 3D/2D views.
The display of the first frame can be slow, but then, thanks to the the \textsc{OpenGL} Display Lists, the Displayer can render more than $60$ frames per second.

Another technique is used to make the display faster for secondary views : after the first draw, an internal screenshot of the view is taken, then, this screenshot is just used to redraw the view.
From time to time the screenshot is updated.  
%However the active view\footnote{The active view is defined the last view that was clicked.} does not use this technique.
However the main view does not use this technique.

The mouse clicks are handled in order to outline (flashing) and print out information of the mouse selected object.

The Figure~\ref{fig:Displayer_Screenshot} shows how the simulated events produced in the appendix~\ref{sec:Complete_Example} looks in the \textsc{Frog} Displayer.
The \textsc{Frog} configuration file used to get this view is available in the same appendix.

\begin{figure}[hb!]
	\centering
	\includegraphics[width=0.70\linewidth]{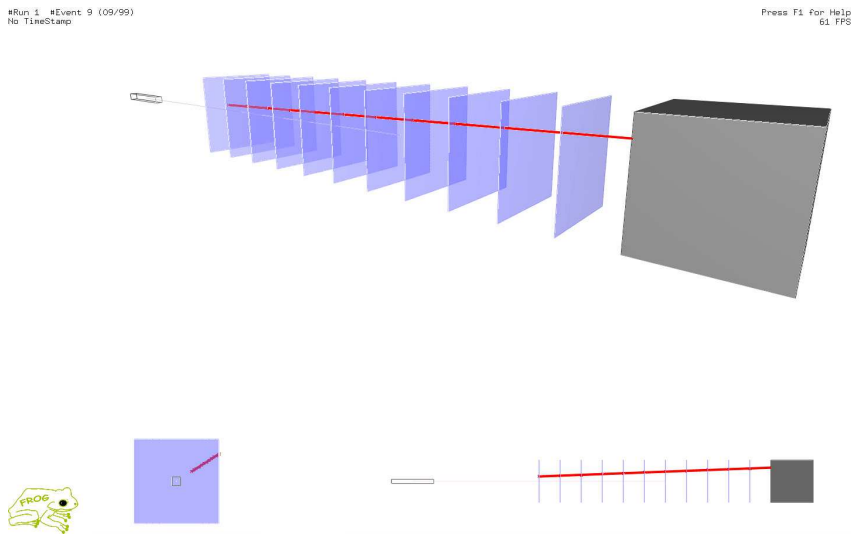}	  
       \caption{Three different \textsc{Frog} Display views: big 3D view (top), 2D longitudinal (bottom right), transversal view (bottom left).  The different geometry parts are well visible : Particle Gun (grey), the Ending Block (grey) and the eleven tracking layers (blue).  The particle track is shown in red.}
  \label{fig:Displayer_Screenshot}	
  \end{figure}

\newpage

\section{Features}
\label{sec:features}

\subsection{Config File}
\label{sec:configFile}

\textsc{Frog} is fully configurable by a set of parameters defined in an ASCII file (\texttt{FROG/config.txt}).  This file contains, amongst other things:
\begin{itemize}
	\item the path to the Input \texttt{.vis} file and/or Input \texttt{.geom} file.
	\item the first event to display.
	\item the objects colour, styles and thresholds.
	\item the views to be used.
\end{itemize}

The parameters may also be defined in other ASCII files that are included by the main configuration file.
Many of these options will be described in the next sections.  

\subsection{Web interface}
The event (\texttt{.vis}) and geometry (\texttt{.geom}) files can be downloaded via internet (world wide web) via the http or ftp protocols.
This can be steered by setting the \textsc{URL} in the configuration file. For example:

\begin{small}
\begin{itemize}
\item \texttt{InputVisFile~= \{http://projects.hepforge.org/frog/tut/02/SimulatedEvents.vis\}};
\item \texttt{InputGeom~~~~= \{http://projects.hepforge.org/frog/tut/02/MyCustomTracker.geom\}};
\end{itemize}
\end{small}

\textsc{Frog} will download automatically the files using the free libcurl \textsc{URL} transfer library: \textsc{LibCURL}~\cite{REF_LIBCURL}.
Some future planned developments that will extend this functionality are discussed in Section~\ref{sec:WebService}.\\
It is possible to update the event file periodically in intervals of time defined in the configuration file:
\begin{itemize}
\item \texttt{updateVisFileTime = 10;~~// File Update Interval in seconds}
\end{itemize}
In this particular case, the event file will be downloaded in a new thread each 10 seconds.  
When negative values are used the file is not updated.
This feature can be useful for visualisation of online events where small \texttt{.vis} files would be produced periodically.
These \texttt{.vis} files could then be put on an http or ftp server and be accessible to every users which have \textsc{Frog} installed on their computer.
Here, the advantage of using multi-threading is clean: the visualisation of the events is not stopped during the download, and the screen would not be frozen.

\subsection{Gzipped files}

Thanks to \textsc{ZLIB}~\cite{REF_ZLIB}, \textsc{Frog} can read gzipped \texttt{.vis/.geom} files automatically, but can also create such files.  It is for example very useful to create automatically \texttt{.vis.gz} files instead of \texttt{.vis} files, in particular when the files are distributed to the users via internet.  The compression rate is around $50\%$ depending on the event data contained in the \texttt{.vis} files.

\subsection{Different Views}

A flexible view system is implemented in \textsc{Frog} in order to increase the number of possible views in the same frame.  
The user can always create the set of views that suits him best only by changing the configuration file.  
Different types of view are available.  The first one is a \textbf{3D view} of the detector.  
The view is defined by a set of parameters in the configuration file:

\begin{itemize}
\item The viewport: region of the screen that should contain the camera view (X,Y,W,H)
\item The Camera Target (X,Y,Z)
\item The Camera Position on a sphere centered on the target (R,$\phi$,$\eta$).
\item The movement of the camera (animate or not)
\item The geometry parts to be displayed in this view
\end{itemize}

The second view type is a \textbf{2D orthogonal projection view}, which is defined also by a set of parameters.  
Most of them are common with the 3D view.  
The 2D views can't be animated.  
A new parameter is used to define the volume to be projected on the screen.
This volume is  a cuboid centered on the camera target, with a surface equal to the screen dimension.  The volume deep is given by the parameter \texttt{Slide\_Depth}.  
So, in the particular case where the 2D camera is aligned with the $z$ axis, the projected objects are contained within an interval in $z$: \texttt{[target.z-Slide\_Depth, target.z+Slide\_Depth]}.
A \textbf{Lego Plot view} of specific detectors (e.g. the calorimeter cells) can also be used.
The code that defines the view classes is fully tunable in order to add easily new view types~\cite{REF_FROGTUT}.

\subsection{Tree menu}

\begin{figure}[hb!]
        \centering
        \includegraphics[width=0.75\linewidth]{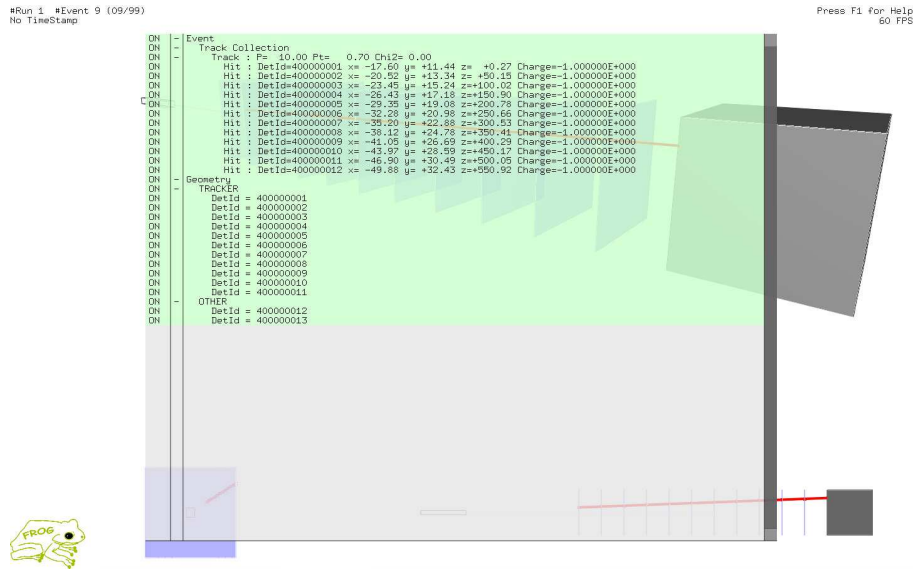}
       \caption{Similar to Fig.~\ref{fig:Displayer_Screenshot} except that the $<$t$>$ menu is opened.  Both the event and geometry structures are visible.  All the lines are green since all the objects are displayed. Hits position and charge are listed as well as the track properties.}
  \label{fig:Features_ScreenshotMenu}
\end{figure}

A menu accessible with key $<$F2$>$ is included in the \textsc{Frog} Displayer.  This menu is very important and simplifies a lot the use of \textsc{Frog}.  The menu is used to specify what content of the \texttt{.vis/.geom} files to display with a  user friendly interface in the form of a structural tree which reflects the structure of the .vis/.geom file (see Fig.~\ref{fig:FROGCore_Chunk_Tree}).  Branches of the tree represent group of objects (detector or event data).  A branch can be opened to see what it contents.  The user can also check and modify the display states of the branch (whether or not visible). A second menu is accessible by the key $<$t$>$.  The $<$t$>$ menu is defined as a semi-transparent view that is superimposed on top of the other views.  It is useful to see in real time the displayed objects changing when states in the menu are modified.  This menu has to be defined in the configuration file as well.  A screenshot showing this menu is presented in Figure~\ref{fig:Features_ScreenshotMenu}.

\subsection{Mouse Interface}
\label{sec:mouseInterface}

In \textsc{Frog}, each object is clickable.  When an object is clicked, some information about the objects are printed on the screen.  As an example, when clicking on a track the momentum, the transverse momentum, the $\chi^2$ and the number of hits are displayed.  Object information to be printed are defined in the class of the selected object.  But the selection routines is an important part of the core of \textsc{Frog}.  (This is completely transparent for the \textsc{Frog} users, even for the users who want to add new objects).

Thanks to the hierarchical structure of the objects it is possible to display the parent of a selected object when clicking on the key $<$1$>$, $<$2$>$, etc.  For example if the selected object is a strip of a tracker module. Hitting $<$1$>$ will display the module of the strip.  Hitting $<$2$>$ will display the grand-mother of the strip: the tracker.  And so on, and so forth.  The key $<$0$>$ and $<$9$>$ are special.  The first just clean up the screen of the parent detectors that have been displayed with $<$1$>$ to $<$9$>$.  While the $<$9$>$ displays the full geometry of the detector.  For complex detector this can dramatically slow down the rendering speed.

\subsection{Online Use}

\textsc{Frog} can be used online to have quickly clues on the quality of data taken by the experiment.  This is transparent to the user, and usage of \textsc{Frog} online and offline only differs by the fact that in the online case, the Producer and the Displayer are used simultaneously.  It is indeed possible to read the .vis file while the Producer is still pushing events into the same file.  The temporary \texttt{.vis} file allows many users to look at the latest events without duplication of heavy and slow physics computation.

It is possible to configure \textsc{Frog} in order to display always the latest event in the file, in order to see events as soon as they are written into the file.
But even in that mode, the user can look at a particular event more deeply just by stopping the event changing (this is done with key $<$s$>$).

\subsection{Production of Pictures}

\textsc{Frog} can easily take screenshot of events and detector in order to produce publishable plots.  There are two fundamental different ways to take screenshots.  The first one, is just to copy all the pixels of the screen and encode them in a particular picture format (e.g. \textsc{PNG}).  The second possibility is to store the information to build the picture.  This is in general called a vectorial picture.  Where objects are stored as a vector of primitives (lines, points, etc).  Some well-known picture formats are using this technique (e.g. \textsc{EPS}, \textsc{PS}, \textsc{PDF}, etc.).

In \textsc{Frog}, two picture libraries are used (so far) in order to work with the two pictures types.  The \textsc{Gl2ps} library~\cite{REF_GL2PS} is used for the vectorial format.  It supports a lot of formats: \textsc{PS}, \textsc{EPS}, \textsc{TEX}, \textsc{PDF}, \textsc{SVG} and \textsc{PGF}.  When a screenshot is taken (with the $<$ENTER$>$ key) in that format (defined in the config file), only the active view is saved.  Also, the output pictures are in general very heavy because of the high precision of that format.  This is the ideal format to use to include \textsc{Frog} pictures in papers.

For the pixelized picture, the \textsc{PNGLIB}~\cite{REF_PNG} and \textsc{ZLIB}~\cite{REF_ZLIB} libraries are used.  This format should be used in all the other contexts because it allows to take screenshots faster and picture file size is lighter.  It is also readable by almost every operating system without special softwares or add-ons.

\subsection{Styles}
\label{sec:Styles}

Each \textsc{Frog} object contains a \texttt{FROG\_Style} as data member, which contains all the style options.  The styles can be changed directly in the configuration file.  Styles of any group of objects can be changed just by using the group EventId or DetId.  Below is an example of the text block to setup styles of a RecoTrack Collection with and EventId (23100001): 

\begin{verbatim}
Id_23100001_Color      = { 1.0 , 0.0 , 0.0 , 1.0 };
Id_23100001_Thickness  = 3.0;    
Id_23100001_Marker     = 0;      
Id_23100001_MarkerSize = 5;      
Id_23100001_ShowDet    = false;  
\end{verbatim}

The {\bf\texttt{Color}} parameter is used to set up the default colour of the track Collection.  
The colour is defined as a vector of \texttt{double} contained between 0 and 1.  With the following syntax \texttt{\{Red, Green, Blue, Alpha\}}.
The {\bf\texttt{Thickness}} parameter defines the thickness of the line used to draw the track.  {\bf\texttt{Marker}} defines the type of the marker used to display the hits associated to the track.

All the available markers can be found in \texttt{FROG/Resources/Marker/*.png}.
Anyone can add markers just by adding new \texttt{.png} picture with same width and height in this directory.  {\bf\texttt{MarkerSize}} is of course used to set up the size of the marker.
Finally {\bf\texttt{ShowDet}} defines if the detector module associated to the hits have to be drawn or not.  In busy events it is recommended to put this parameter on \texttt{false}.\\
If the style of an object is not defined, the style of his mother is used.  If no parents have styles, unpredictable style is used.

\newpage

\section{Examples of Applications}
\label{sec:Applications}

\textsc{Frog} is already used in different experiments and environments, some of them are shown in this section.  The two examples described in this section show that \textsc{Frog} can be used in many different applications from the experiment to the simulation environement: \textsc{GasTOF} is a very small detector prototype~\cite{REF_GASTOF}.  The second example, \textsc{Delphes} is not a detector but a framework for the fast simulation of the response of a generic detector in high energy physics~\cite{REF_DELPHES}.  %Finally CMS is one of the four LHC experiment, it has a very large and complex geometry that is in general hard to render properly.

\subsection{\textsc{GasTOF}: The Ultra-Fast Gas Time-of-Flight Detector}

\textsc{GasTOF}~\cite{REF_GASTOF}\cite{REF_GASTOF2} detectors are Cherenkov gas detectors that will be located at $420~\textrm{m}$ from the CMS~\cite{REF_CMS} and ATLAS~\cite{REF_ATLAS} interaction point (IP), as part of the FP420 project~\cite{REF_FP420} for the LHC~\cite{REF_LHC}.  
The aim of these detectors is to reduce backgrounds due to accidental coincidence of events detected in the central detectors and in the FP420 detectors on each side of the IP.
To achieve that, the event vertex $z$-coordinate measured by the central detectors is compared to the vertex reconstructed by measuring the time difference of forward proton arrivals to the \textsc{GasTOF} detectors on two sides of the given IP.
It requires a very precise measurement of the proton time of flight ($\delta t \sim 10$-$20$ ps).
Cherenkov photons produced by high energy protons traversing gas medium are reflected by a mirror onto a very fast photomultiplier. 
In gases, thanks to small refractive index, these photons are emitted at very small angles.

\subsubsection{Display of the Geometry}

\begin{figure}[ht!]
	\centering
	\includegraphics[width=1.0\linewidth]{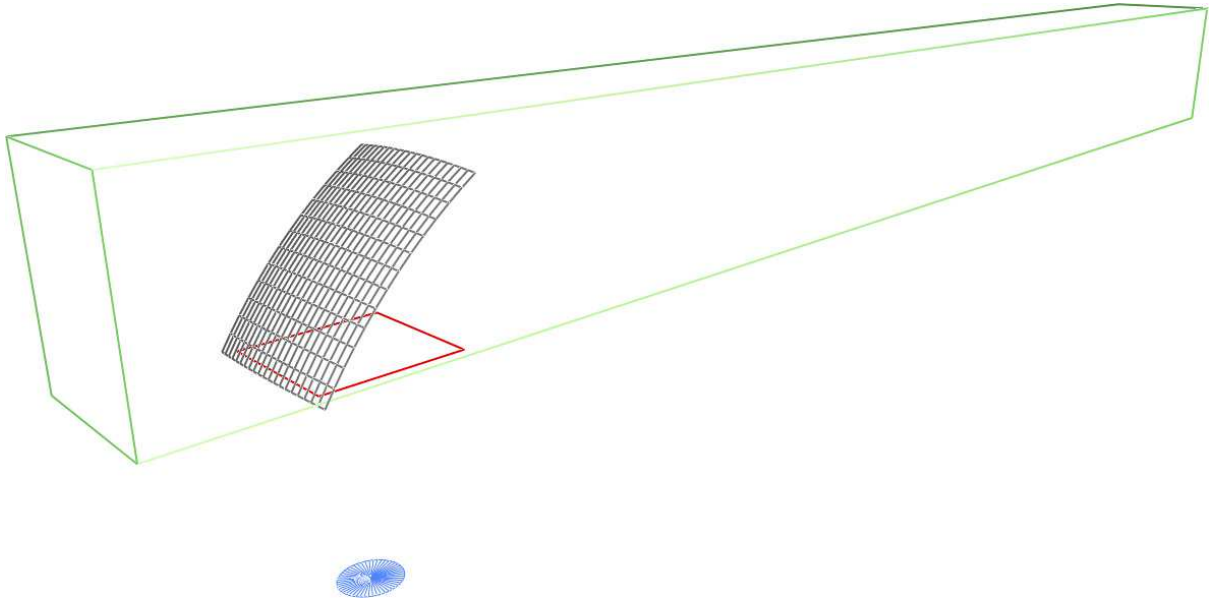}	  
	\caption{ A 3D View of the full \textsc{GasTOF} geometry (in wireframe). The gaseous Cherenkov detector is represented by the green box.  Inside the box, there is a toroid mirror (grey).  A window (red) is present in the box to allow photons focused by the mirror to reach the \textsc{GasTOF} photomultiplier.  The sensitive zone of the PMT photocathode is represented by the blue circle.  Only the sensitive zone of the photomultiplier is drawn since the other parts do not affect the photon detection.}	
	\label{fig:Applications_Gastof_Geom}
\end{figure}

In the simulation, \textsc{GasTOF} is modelled as a finite cuboid volume filled with gas ($C_4F_{10}$).  Thanks to a toroid mirror placed inside the box, the Cherenkov photons are reflected and focused onto a small photocathode, leaving the box by a "window".  All the elements are shown on Figure~\ref{fig:Applications_Gastof_Geom}, for the $31~\textrm{cm}$ long \textsc{GasTOF} prototype used for the June 2008 test beam at CERN.
\vspace{1cm}

\newpage

\subsubsection{Display of the Events}

\begin{figure}[ht!]
	\vspace{0.5cm}
	\begin{minipage}[htbp]{0.49\linewidth}
		\centering
		\includegraphics[width=1.0\linewidth]{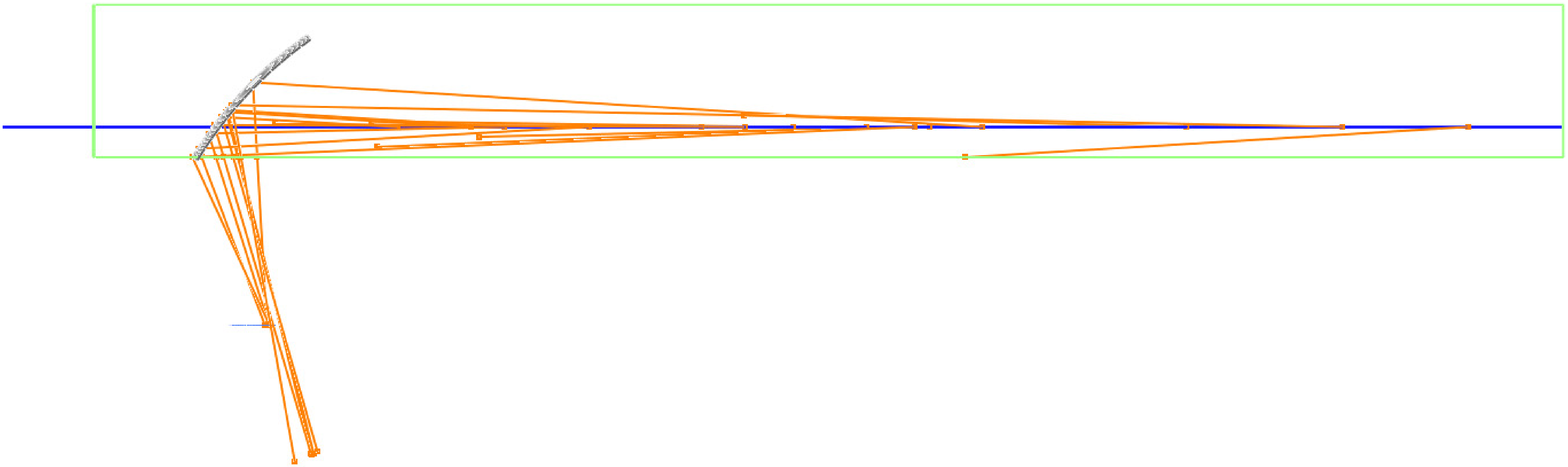}	  
	\end{minipage}
	\hfill
	\begin{minipage}[htbp]{0.49\linewidth}	
		\centering
		\includegraphics[width=1.0\linewidth]{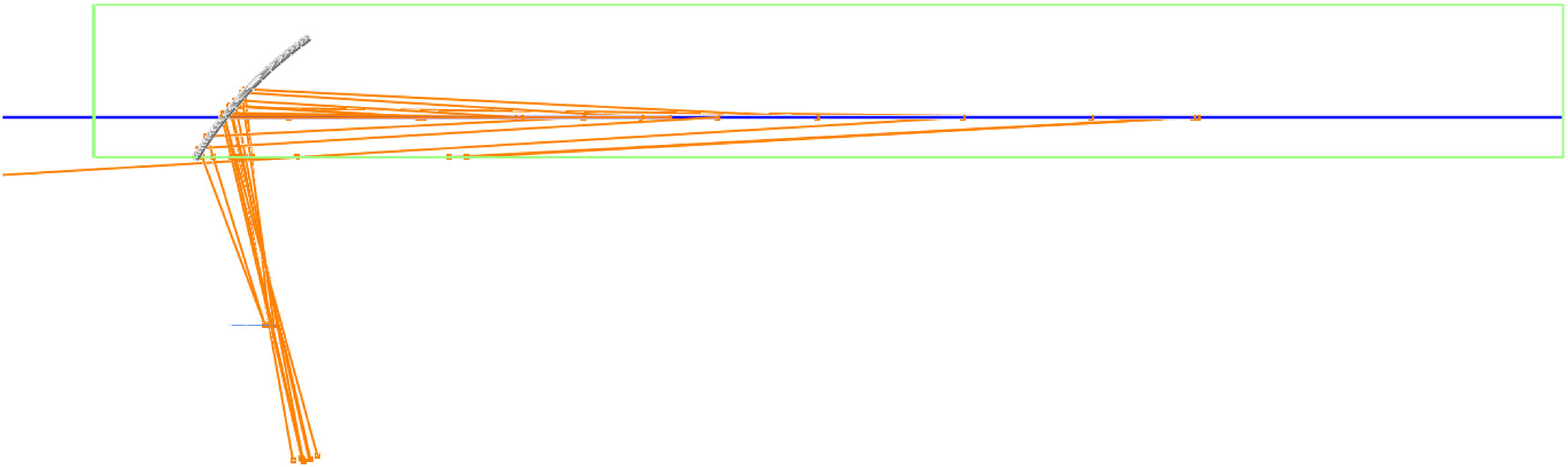}	  
	\end{minipage}
	\vspace{0.5cm}
	\begin{minipage}[htbp]{0.49\linewidth}
		\centering
		\includegraphics[width=1.0\linewidth]{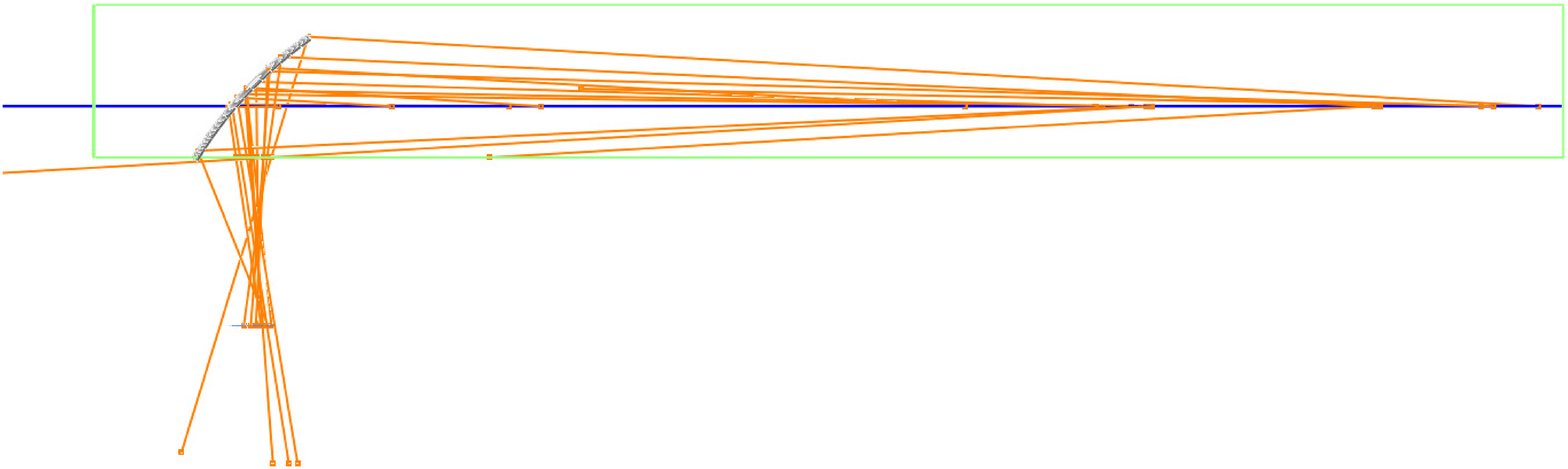}	  
	\end{minipage}
	\hfill
	\begin{minipage}[htbp]{0.49\linewidth}	
		\centering
		\includegraphics[width=1.0\linewidth]{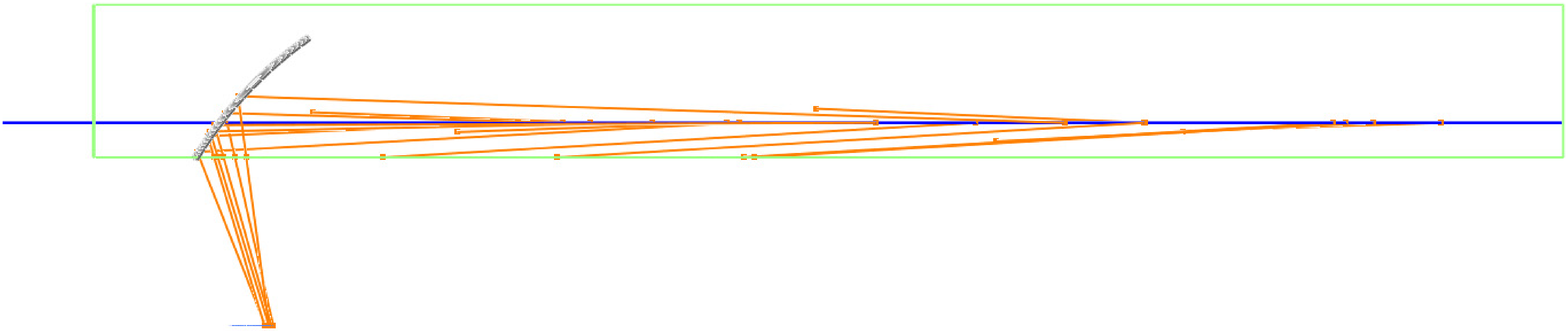}	  
	\end{minipage}
	\vspace{0.5cm}	
	\begin{minipage}[htbp]{0.49\linewidth}
		\centering
		\includegraphics[width=1.0\linewidth]{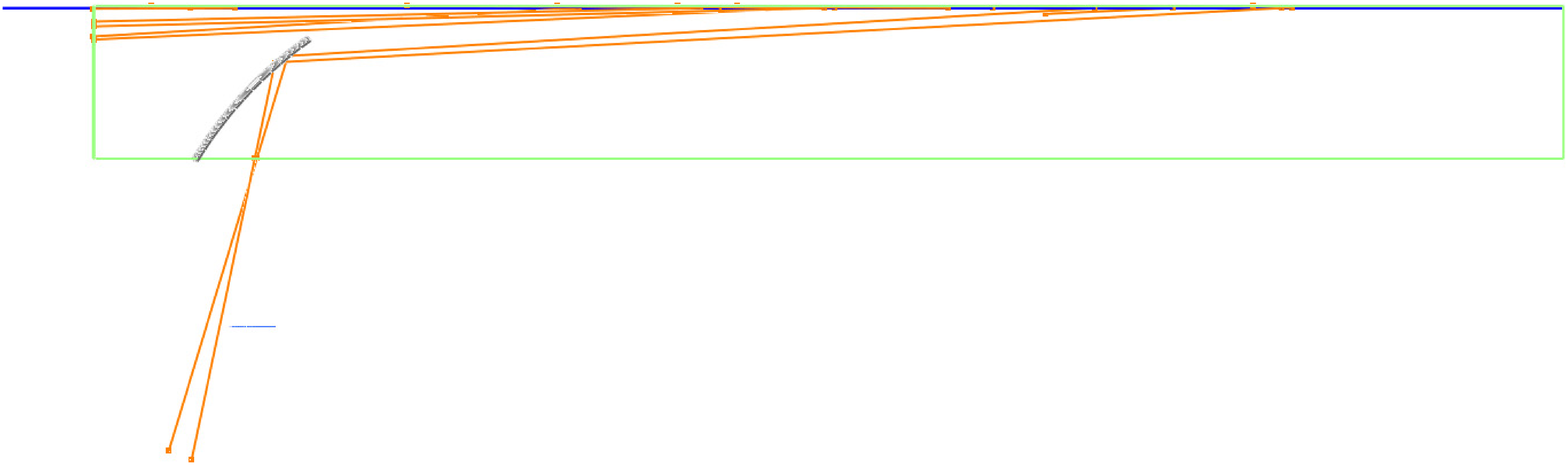}	  
	\end{minipage}
	\hfill
	\begin{minipage}[htbp]{0.49\linewidth}	
		\centering
		\includegraphics[width=1.0\linewidth]{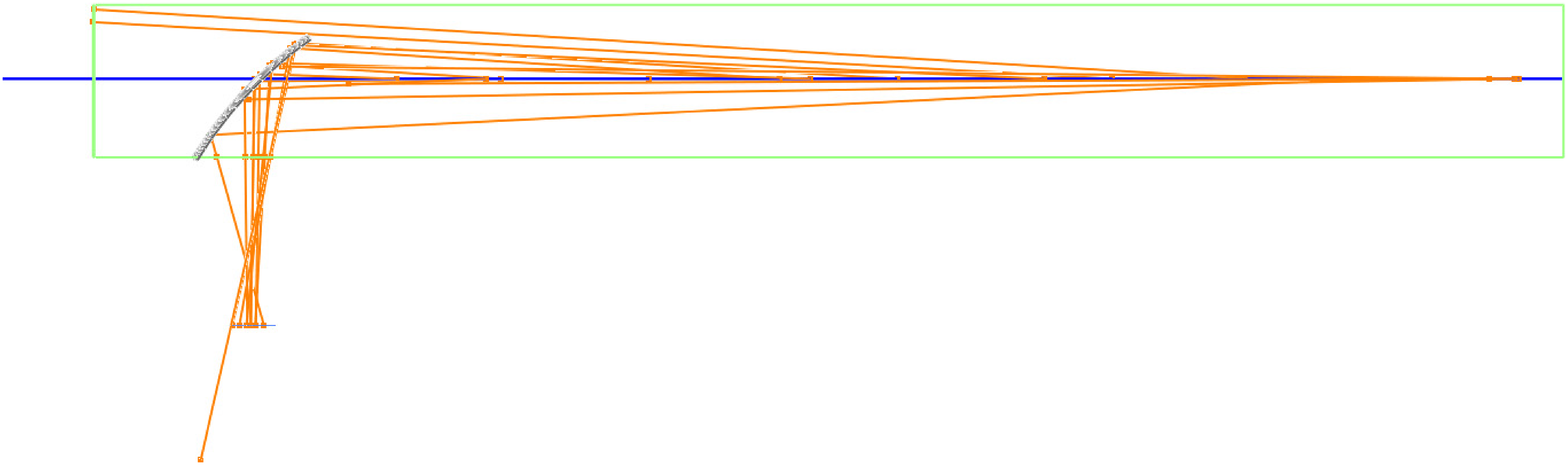}	  
	\end{minipage}
	\caption{ 2D views of six different \textsc{GasTOF} events: The proton trajectory (going from right to left in that view) is represented by the blue line.  While photon trajectories (ray) are shown in yellow.  
The Cherenkov photon production is well visible along the proton path.  The majority of produced photons are reflected by the curved mirror and focused on the photo-cathode.  
Note that some photons are not reflected on the mirror and than some others miss the PMT.  }	
	\label{fig:Applications_Gastof_Event_2D}
\end{figure}

\begin{figure}[ht!]
%	\begin{minipage}[htbp]{0.7\linewidth}
		\centering
		\includegraphics[width=0.95\linewidth]{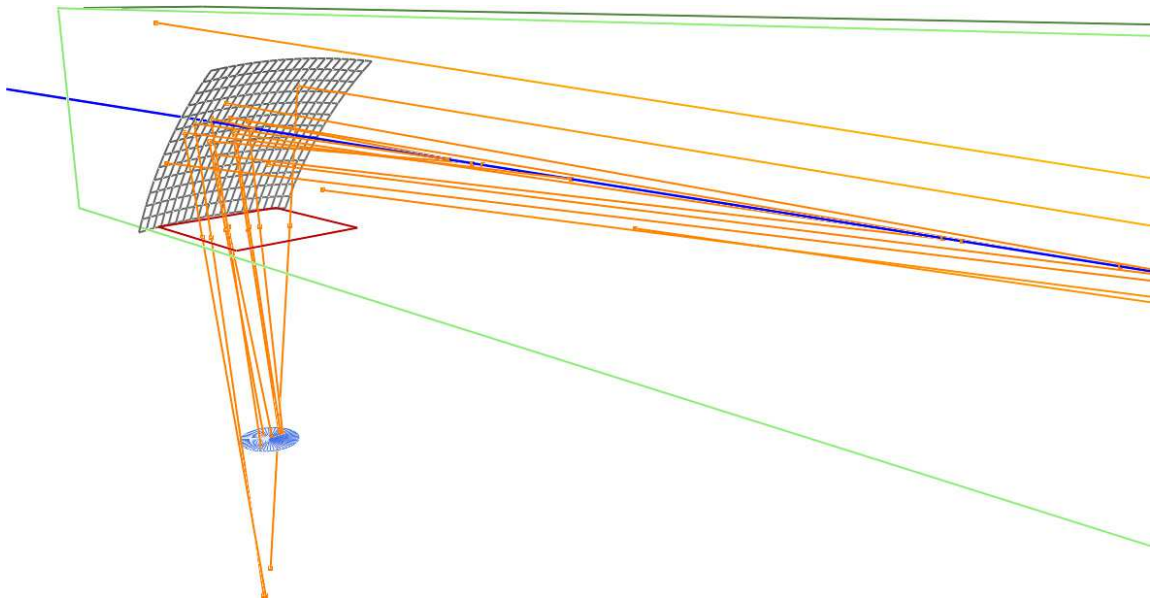}	  
%	\end{minipage}
%	\hfill
%	\begin{minipage}[htbp]{0.49\linewidth}	
%		\centering
%		\includegraphics[width=1.0\linewidth]{Pictures/Applications_Gastof_Event_3DB.eps}	  
%	\end{minipage}
	\caption{ 3D view zoomed on the \textsc{GasTOF} mirror. Legend and colour codes have already been described in Figure~\ref{fig:Applications_Gastof_Event_2D}.  }	
	\label{fig:Applications_Gastof_Event_3D}
\end{figure}

The figures below show the simulated \textsc{GasTOF} events comming from the ray-tracing done by the \textsc{GasTOF} simulator~\cite{REF_GASTOF_SIM}.
The Figure~\ref{fig:Applications_Gastof_Event_2D} shows 2D Y-Z projections of six different events while the Figure~\ref{fig:Applications_Gastof_Event_3D} shows a 3D view of an other event.  
In both cases, the Cherenkov photons paths (ray) are drawn in yellow.

\vspace{0.5cm}

\clearpage

\subsection{\textsc{Delphes}:\newline a Framework for the Fast Simulation of a General Purpose Collider Experiment}
%\subsection*{\textsc{Delphes}:\newline a Framework for the Fast Simulation of a General Purpose Collider Experiment}
%\addcontentsline{toc}{subsection}{\textsc{Delphes}}

Knowing whether theoretical predictions are visible and measurable in a high energy experiment is always delicate, due to the
complexity of the related detectors, their data acquisition chain and their operating software. The \textsc{Delphes} framework~\cite{REF_DELPHES} has been introduced for the fast and realistic simulation of a general purpose experiment, like CMS or ATLAS at the LHC. The simulation includes the usual components of such detector as well as possible very forward detectors arranged along the beamline.

The framework is interfaced to standard file formats (e.g. Les Houches Event File) and outputs observable analysis data objects, like missing transverse energy and collections of electrons or jets.
The simulation of detector response takes into account the detector resolution, by smearing the kinematic properties of the particle.
Usual reconstruction algorithms are applied for complex objects, like the missing transverse energy or the jets originating from $b$ quarks or $\tau$ leptons. 

A simplified preselection can also be applied on processed data for trigger emulation. Detection of very forward scattered particles relies on the transport in beamlines. Finally, the visualisation of the collision final states is possible via a dedicated Producer interfacing \textsc{Frog} to \textsc{Delphes}.

A fast simulation can be used to obtain realistic observables and fast estimates of signal and background rates for specific channels.  Starting from generator-level information, the package provides reconstructed jets, isolated leptons, photons, reconstructed charged tracks, calorimeter towers and the expected transverse missing energy. 

\begin{figure}[hb!]
	\begin{minipage}[htbp]{0.49\linewidth}
		\centering		
		\includegraphics[width=1.0\linewidth]{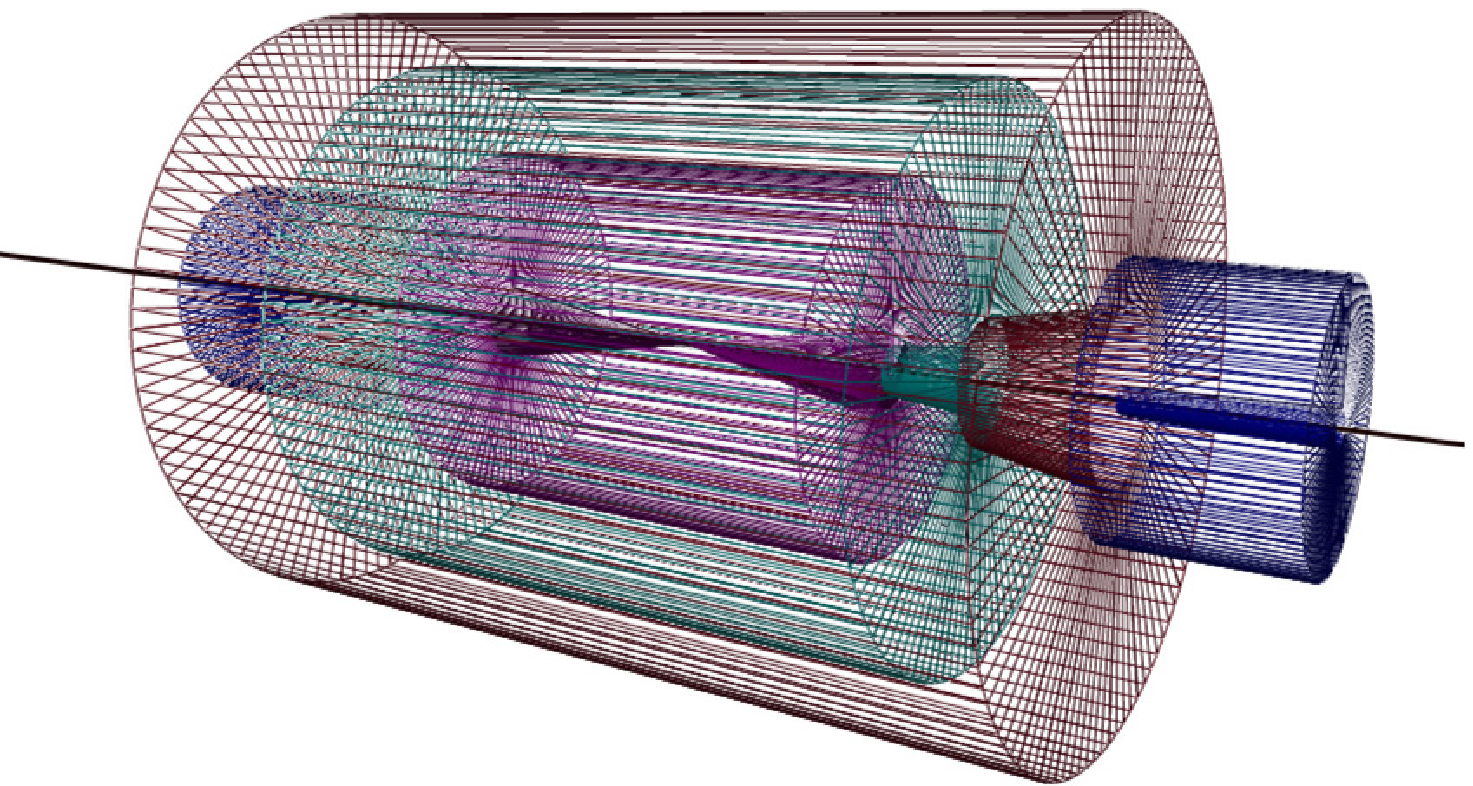}	  
	\end{minipage}
	\hfill
	\begin{minipage}[htbp]{0.39\linewidth}	
		\centering
		\includegraphics[width=1.0\linewidth]{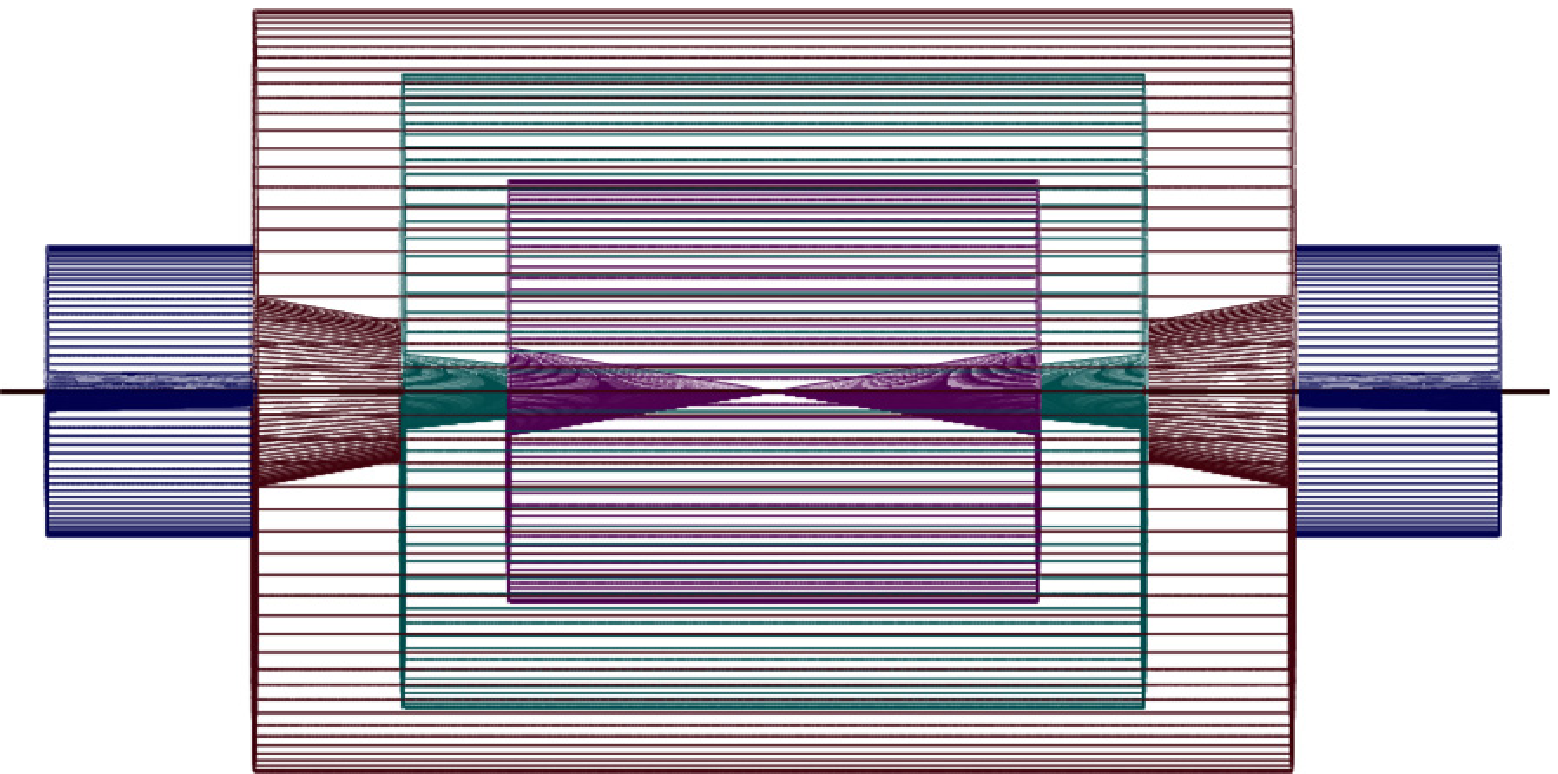}	  
	\end{minipage}
	\caption{Left: Layout of the generic detector geometry assumed in \textsc{Delphes}. The innermost layer, close to the interaction point, is a central tracking system (pink). It is surrounded by a central calorimeter volume (green) with both electromagnetic and hadronic sections. The outer layer of the central system (red) consist of a muon system. In addition, two end-cap calorimeters (blue) extend the pseudorapidity coverage of the central detector. The actual detector granularity and extension is defined in the user-configuration card. The detector is assumed to be strictly symmetric around the beam axis (black line). Additional forward detectors are not depicted.  Right: Profile of the layout assumed in \textsc{Delphes}. The extension of the various subdetectors, as defined in Tab.~\ref{tab:DELPHES_defEta}, are clearly visible.}
	\label{fig:DELPHES_GenDet}	
\end{figure}

\begin{figure}[ht!]
\begin{center}
\includegraphics[width=0.8\columnwidth]{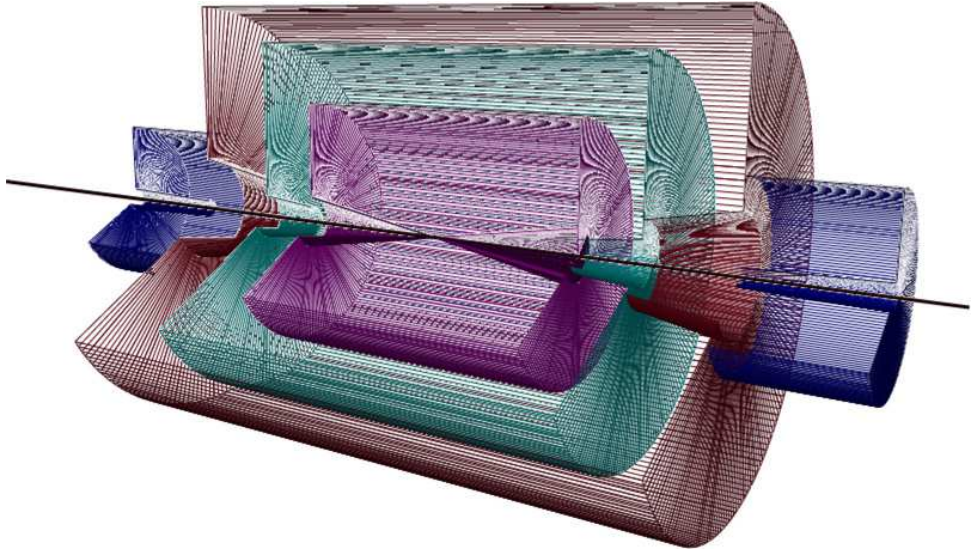}
\caption{Layout of the generic detector geometry assumed in \textsc{Delphes}. Open 3D-view of the detector with solid volumes. Same colour codes as for Fig.~\ref{fig:DELPHES_GenDet} are applied. Additional forward detectors are not depicted.}
\label{fig:DELPHES_GenDet2}
\end{center}
\end{figure}

The overall layout of the general purpose detector simulated by \textsc{Delphes} is shown in Fig.~\ref{fig:DELPHES_GenDet} and~\ref{fig:DELPHES_GenDet2}. A central tracking system surrounded by an electromagnetic and a hadron calorimeters. The muon system encloses the detector volume. A forward calorimeter ensure a larger geometric coverage for the measurement of the missing transverse energy. The fast simulation of the detector response takes into account geometrical acceptance of sub-detectors and their finite energy resolution. All detectors are assumed to be symmetric with respect to the beam axis. The configuration of the subsystems used in these examples is summarised in Table~\ref{tab:DELPHES_defEta}.
\vspace{0.5cm}

\begin{table}[hb!]
	\vspace{0.5cm}
	\centering
	\begin{tabular}[!h]{llc}
	\hline
System & \multicolumn{2}{l}{Extension in pseudorapidity} \\
\hline
Tracking     &  & $0 \leq |\eta| \leq 2.5$\\
Calorimeters & central & $0 \leq |\eta| \leq 3.0$\\
             & forward & $3.0 \leq |\eta| \leq 5.0$\\
Muon system  &  & $0 \leq |\eta| \leq 2.4$ \\\hline
	\end{tabular}
	\caption{Configuration of the subsystems used in the examples presented in Fig.~\ref{fig:DELPHES_GenDet} and~\ref{fig:DELPHES_GenDet2} }
	\label{tab:DELPHES_defEta}
\end{table}

\newpage

\begin{figure}[hb!]
\begin{center}
\includegraphics[width=\columnwidth]{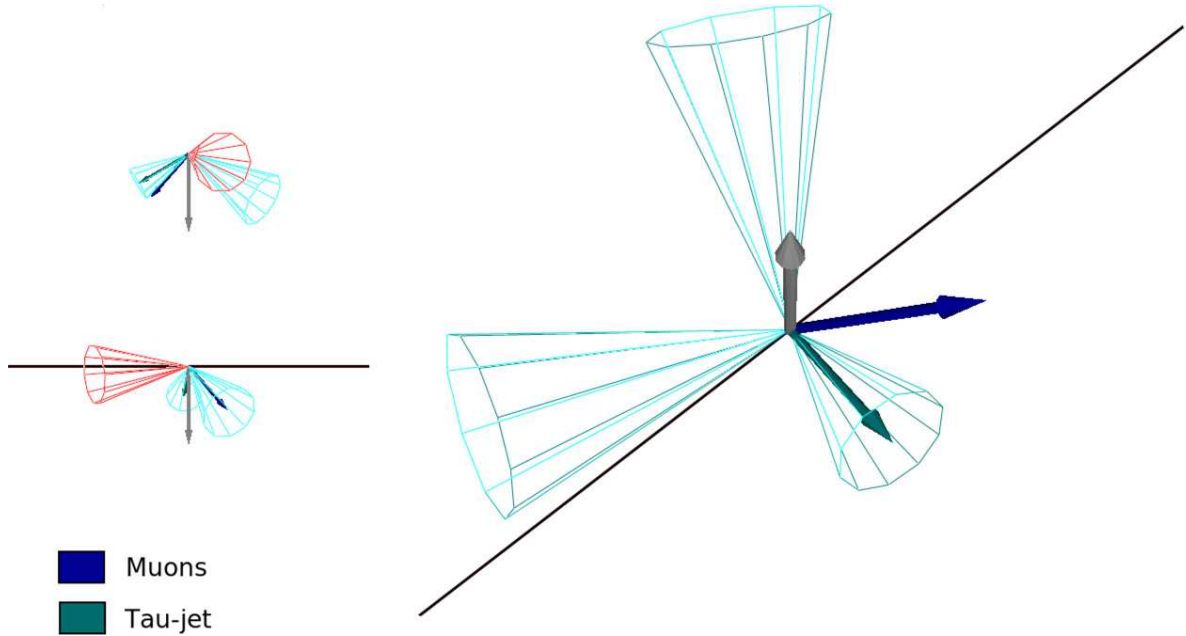}
\caption{Example of $pp(\gamma p \rightarrow Wt)pY$ event. One $W$ boson decays into a $\mu \ \nu_\mu$ pair and the second one into a $\tau \ \nu_\tau$ pair. The surviving proton leaves a forward hemisphere with no hadronic activity. The isolated muon is shown as the blue vector. The $\tau$-jet is the cone around the green vector, while the reconstructed missing energy is shown in gray. One jet is visible in one forward region, along the beamline axis, opposite to the direction of the escaping proton.}
\label{fig:DELPHES_wt}
\end{center}
\end{figure}

As an illustration, an associated photoproduction of a $W$ boson and a $t$ quark is shown in Fig.~\ref{fig:DELPHES_wt}. This corresponds to a $pp \rightarrow Wt \ +  \ p  \ + \ X$ process, where the $Wt$ couple is induced by an incoming photon emitted by one interacting proton. This leading proton survives from the photon emission and subsequently from the $pp$ interaction, and is present in the final state. The experimental signature is a lack of hadronic activity in one forward hemisphere, where the surviving proton escapes. The $t$ quark decays into a $W$ and a $b$. Both $W$ bosons decay into leptons ($W \rightarrow \mu \nu_\mu$ and $W \rightarrow \tau \nu_\tau$). 

\clearpage

\section{Future Developments}
\label{sec:pespectives}

Since \textsc{Frog} is a very flexible tool that can be considered more as a toolkit than as static software, many perspectives are possible for the evolution of the software.
By design, \textsc{Frog} can grow very quickly because all the displayable objects are defined by their own C++ class inheriting from the same base class.
Therefore, users can add new needed objects themselves in the software.  Moreover, they can also add new types of view.  

All the basic tools are ready to be used in a larger scale.  Some experiments already started to intensively use the software and numerous feedbacks from this large community of users helped to track bugs and improve the tool.

The short term perspectives consist in the improvement of the tree menu in order to allow the user to change styles, colours, cuts, etc... directly from the menu.  By this way, the users will not have to edit any more the \textsc{Frog} configuration file by themselves.

An other improvement which will be implemented is the possibility to filter events contained in the \texttt{.vis} file directly from the Displayer in order to easilly find events satisfying simple criteria.

On the long term perspectives, the possibility to turn \textsc{Frog} into a complete plug-in system will be investigated. If the code implementation is simplified for the user and performances are at least equal, the plugin system will be adopted. 
 
An other long term perspective is to improve the image quality of \textsc{Frog} by improving the 3D perspectives and effects.  This can easily be done by adding lighting/shaddowing effects in the 3D scene.  This makes in general the feeling of the perspective much better without adding too much complication in the code.  This effect has also the advantage to be supported by any \textsc{OpenGL} VGA card.

Of course, users' requests will also be included in the next releases.

\subsection{\textsc{Frog} Web Service}
\label{sec:WebService}

In collaboration with the Caltech CMS group~\cite{REF_CALTECH}, a pilot project has been started.  Its goal is to offer a Web service that allows users to easily produce the FROG visualisation files (\texttt{.vis}, \texttt{.geom}) from CMSSW~\cite{REF_CMSSW} datasets.
This service will offer to users, who do not necessarily have the full CMS software installed locally, a mean to remotely create and then download the Frog files. Since these \texttt{.vis} and \texttt{.geom} files are relatively small ($\sim$14\rm{Mb}/200 events) and the CMSSW processing time to create them is minimal, requests for \textsc{Frog} files from average sized datasets can be processed rapidly and downloaded quickly using standard broadband network connections.

In the first phase of this project, the \textsc{Frog} Web Service will allow the selection of a CMSSW dataset (containing either simulated or real events) using an interface to DBS, a requested range of runs and/or event numbers, and a choice of which version of CMSSW is to be used to process the data into \textsc{Frog} files. In addition, a browsable and searchable set of previously created \textsc{Frog} files will be provided to the user.
In a second phase, the Web Service would be enhanced to allow customized CMSSW configuration files to be uploaded to the server, and to allow the automatic creation of standard \textsc{Frog} projections/displays, which would be stored as graphics files on the server and made available for download via a photo gallery style interface.
\newpage
\section{Summary}

The goal of the \textsc{Frog} project was to create a new event display able to quickly display a significative fraction of the event data, in a high energy physics experiment. As a high rendering speed is desirable, the \textsc{Frog} software has been splitted into two parts: the Producer and the Displayer.  In addition to the speed enhancement, this structure coupled with a "home-made" file format brings many interesting features for free, like the portability of the framework on several operating systems, the independence with respect to the experiment or environment, the possibility to share easily visualisation data, etc. Another important application of \textsc{Frog} is the online experiment data display, as the Producer and the Displayer can work simultaneously.

In addition of the previous features, \textsc{Frog} is based on an object oriented design.  Therefore, it can display almost any imaginable object without any sensitive lost of speed.  It can work on almost any recent computer with one of the following Operating Systems: Linux, MacOs or Windows.  The size of the full \textsc{Frog} package, including the source code, is less than $4~\textrm{Mb}$.  It is really suitable for outreach applications since it can be freely distributed and installed in only few clicks. Moreover, latest available data can be automatically downloaded from the internet by the \textsc{Frog} Displayer.  One of the main goal of the Displayer was to allow a large number of users to visualise event data in quasi-real time.

In conclusion, \textsc{Frog} can be used in many applications: it is highly tunable and allows a large number of users to visualise event data.  The software is already used by several physicist communities for outreach and detector debugging.

\vspace{2cm}

\section*{Acknowledgements}

First of all we would like to thanks the many people that helped us in the writting of this paper, in particular Giacomo Bruno, Dorian Kcira and Xavier Rouby.  
Furthermore, we thanks the different collaboration that have contributed to this note : \textsc{GasTOF} , Delphes and the Caltech CMS Group.

\newpage
\section{Appendices}
%section*{Appendices}
%\addcontentsline{toc}{section}{Appendices}
\appendix

\subsection*{Appendix A: Base Classes}
\addcontentsline{toc}{subsection}{Appendix A: Base Classes}
\label{sec:base_classes}

%\section{Base Classes}

\begin{figure}[hb!]
	\centering
	\includegraphics[width=1.0\linewidth]{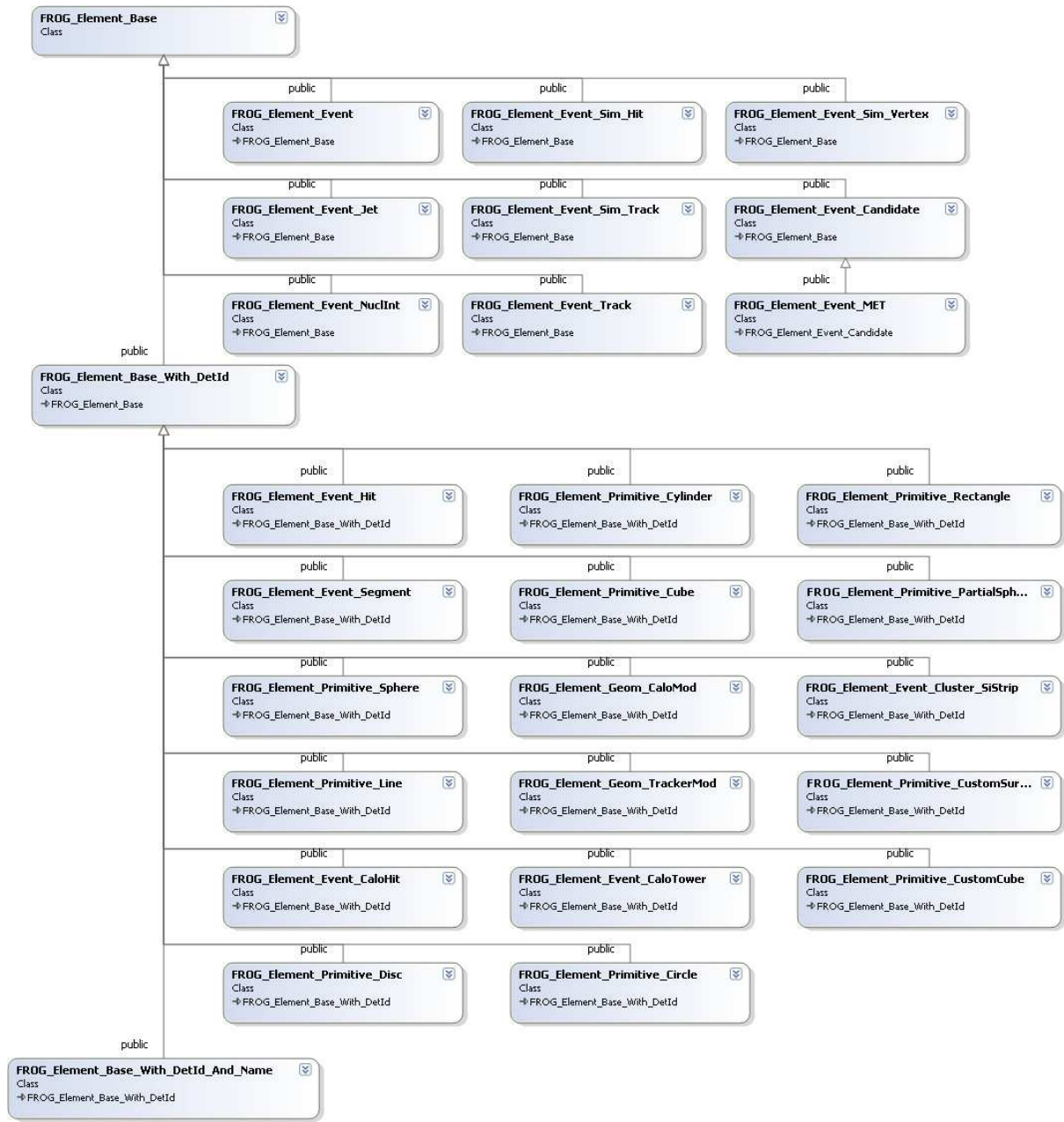}	  
	\caption{The \textsc{Frog} class hierarchy of the displayable objects. All the displayable object classes inherit from \texttt{FROG\_Element\_Base}.  The three classes on the extreme left of this diagram are generally used as objects containers (branch).}	
	\label{fig:CORE_ClassDiagram1}
\end{figure}

\begin{figure}[ht!]
	\centering
	\includegraphics[width=1.0\linewidth]{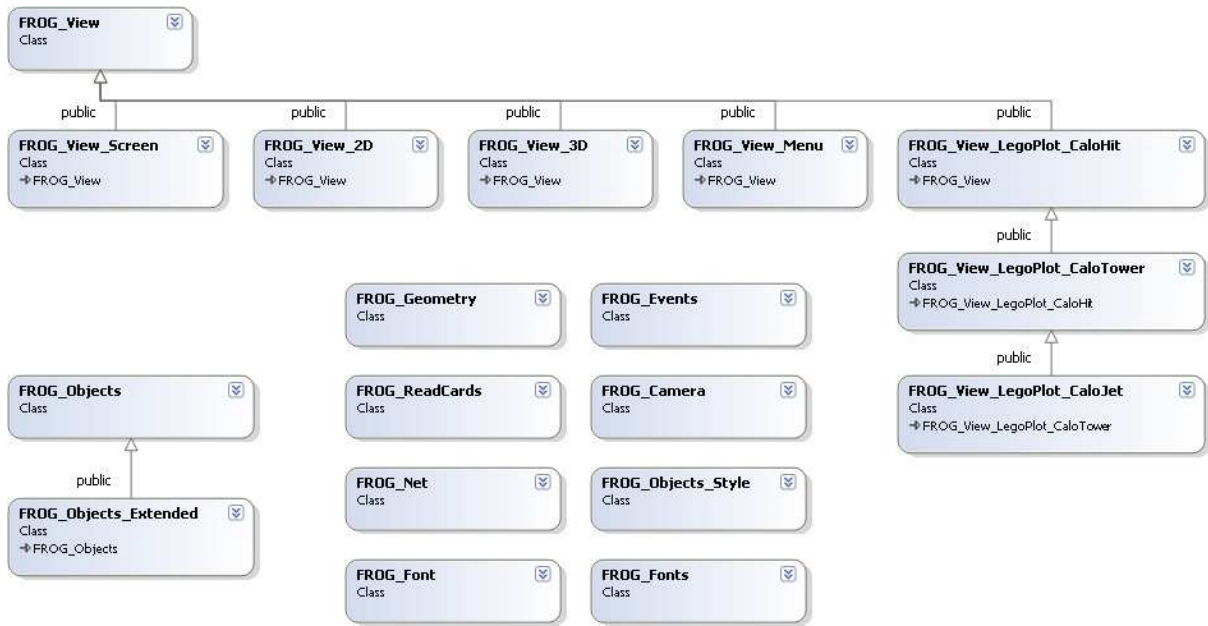}	  
	\caption{The \textsc{Frog} class hierarchy of the other objects like the different views.}	
	\label{fig:CORE_ClassDiagram2}
\end{figure}

This appendix details some of the main classes used in \textsc{Frog}.
The class diagram on Figure~\ref{fig:CORE_ClassDiagram1} shows that all the displayable object classes inherit from \texttt{FROG\_Element\_Base} and/or \texttt{FROG\_Element\_Base\_With\_DetId}.
Clues on how these classes work are given for information, but users have not to modify any of the class described here, even if the user want to use the code for its experiment.
The class \texttt{FROG\_Element\_Primitive\_Line} is used to give a short description of which methods a class associated to a displayable object has to contain. 
The hierarchy of the other \textsc{Frog} classes is shown in Figure~\ref{fig:CORE_ClassDiagram2}.\newline

\newpage

\subsubsection*{FROG\_Element\_Base}
\label{sec:FROGElementBase}

All displayable \textsc{Frog} objects are represented by a class that inherit from \texttt{FROG\_Element\_Base}.  
This base class contains several methods used to read/write the \textit{data}.
These classes are used both by the Displayer and the Producer.

\paragraph*{\textit{Data Members}}
~~\\[-8mm]
\begin{small}
\begin{verbatim}
unsigned short                     type_;
unsigned int                       size_;
unsigned char*                     data_;
std::vector <FROG_Element_Base*>   daughters_;
FROG_Element_Base*                 mother_;
unsigned char                      display_;
unsigned int                       display_list_;
FROG_Objects_Style*                style_;
\end{verbatim}
\end{small}

The three first variables define the chunk: the \textit{Id}, the \textit{size} and the \textit{data}, they are managed directly by the \texttt{read()} and \texttt{write()} methods.
The \texttt{daughters\_} and \texttt{mother\_} are necessary to navigate in the object's tree.  
The next variables are used to optimise the display.  \texttt{display\_list\_} stores a pointer to the object display list~\cite{REF_DLIST}\footnote{A display list is a compiled version of the display code stored in the graphic card memory.  It is used to improve the rendering speed.}.
Finally, \texttt{style\_} contains the aesthetic properties of the object (e.g. colour, thickness, etc.) and the physical cuts ($P_T^{min}$, $E_T^{min}$, etc.) to be applied.

\paragraph*{\textit{Default Constructor}}
~~\\%[-3mm]
The default \texttt{constructor} of \texttt{FROG\_Element\_Base} initialises the data to a null pointer, the \texttt{type\_} and the \texttt{size\_}.

\paragraph*{\textit{Writing method}}
~~\\%[-3mm]
By design, the only data contained in \texttt{FROG\_Element\_Base object} are sub-chunks, it is a branch\footnote{if the chunk \textit{Id} is equal to \texttt{C\_PRIMARY} the branch is actually the root.}. 
So, the \texttt{write} method of this class has only to store the objects contained in the \texttt{daughters\_} variable into the \texttt{data\_} data member.  
%In the case where the object has no mother, the data are stored in a \texttt{.vis} or \texttt{.geom} file.  In the other case , the data are stored in memory and will be actually saved on disk by the mother object.

\begin{small}
\begin{verbatim}
// Fill data_ and size_ with data
virtual void write(){
   std::vector< std::pair<unsigned int, unsigned int > >* blockOfDaughters;
      
   blockOfDaughters = write_init();      
   if(blockOfDaughters==NULL)return;      
   write_daughters(blockOfDaughters);
}
\end{verbatim}
\end{small}

The \texttt{write\_init()} method set up the object and its daughters for storing: it allocates memory, it looks for the optimal storing strategy, etc.
The outputs of this function are groups of objects that have to be stored in a same chunk.
This output is used by \texttt{write\_daughters()} which fill the variable \texttt{data\_} that will be used later to produce the \texttt{.vis} or \texttt{.geom} file.
%to writes \texttt{data\_} the daughter's \textit{data} in memory.

\paragraph*{\textit{Reading method}}
~~\\
The \texttt{read()} is not really implemented in this class.% since do not contains \textit{data}.

\paragraph*{\textit{Initialisation method}}
~~\\
\texttt{FROG\_Objects} is a global structure containing useful variables like the events collection, the geometry, the parameters read from the configuration files, etc.
The \texttt{init(\ldots)} method is used to share this pointer to the entire tree and to reset the object styles and cuts to its initial state using configuration parameters.

\paragraph*{\textit{Display}}
~~\\
The display is done by \texttt{virtual~void~display(bool~UseDisplayList=true,~float*~ color=NULL)}. 
The display lists~\cite{REF_DLIST} improve rendering performances, but in some particular cases, they can not be used, this is the reason of the first argument of \texttt{display}.
The second argument is a pointer on a \texttt{float[4]} that contains the four components of the colour to use for the object to be drawn.  
When this argument is \texttt{NULL}, the colour used is taken from the $style\_$ variable.  

In the case of a \texttt{FROG\_Element\_Base} the display function just call iteratively the display of its daughters and put the result in a display list.

\paragraph*{FROG\_Element\_Base\_With\_DetId}
~~\\

In large physics experiment, it is convenient to identify each subset of detectors/data; this Id is generally simply an integer.
For this reason, the \texttt{FROG\_Element\_Base} class has been extended to a class called \texttt{FROG\_Element\_Base\_With\_DetId}\footnote{This class has itself been extended to a \texttt{FROG\_Element\_Base\_With\_DetId\_And\_Name} which contains a string in addition of the integer.}.
It is a perfect clone of the base class, except that it contains an extra variable \texttt{detId\_} of type \texttt{unsigned int}.

\paragraph*{\textit{Writing method}}
~~\\
This new data members is saved by the \texttt{write()} method of the extended class.  

\begin{small}
\begin{verbatim}
// Fill data_ and size_ with data
virtual void write(){
   std::vector< std::pair<unsigned int, unsigned int > >* blockOfDaughters;
      
   blockOfDaughters = write_init();
   if(blockOfDaughters==NULL)return;
   data_ = FillBuffer(data_, &detId_, sizeof(int)); // write the detId_
   write_daughters(blockOfDaughters);
}
\end{verbatim}
\end{small}

This extended \texttt{write()} method is completely identical to the \texttt{write()}  of the base class except that it contains an extra line that copy the value of \texttt{detId\_} in \texttt{data\_}.
Similarly, the \texttt{FROG\_Element\_Base\_With\_DetId} constructor has to be modified to properly initialise the \texttt{detId\_}, when reading from a file (\texttt{.geom} or \texttt{.vis}). 

\paragraph*{\textit{Reading method}}
~~\\[-8mm]
\begin{small}
\begin{verbatim}
FROG_Element_Base_With_DetId(unsigned short type, FILE* pFile):
FROG_Element_Base(type){
   size_ = sizeOf();
   fread(&detId_ ,sizeof(detId_),1,pFile);		
}
\end{verbatim}
\end{small}

\paragraph*{\textit{\texttt{sizeOf()} method}}
~~\\%[-3mm]
The \textit{size} of the chunk has changed, so the \texttt{sizeOf()} method has to be updated:
\begin{small}
\begin{verbatim}
 static unsigned int sizeOf(){
    return 6 + sizeof(unsigned int);
 }
\end{verbatim}
\end{small}

The \texttt{sizeOf} method always returns the number of bytes needed to store an object of this class, it is always $6$ plus the sum of the \texttt{sizeof()} of all variables stored by this object.
This exetend class contains new functions related to \texttt{detId\_}:  a method to get the \texttt{detId}, a method to look for an object with a given \texttt{detId}, etc.

\paragraph*{The Reading Procedure}
~~\\

The reading procedure is handled by a global \texttt{Read} method, which scans iteratively all the chunks contained in a file until the number of bytes read exceed the expected number of bytes to read.  
A pointer to the mother object is given as an argument of this function, this allows to attach objects created from sub-chunks as daughters of a mother.
When a chunk with a known chunk Id is found, a new object of the proper type is created as a daughter of the mother chunk. 
If this chunk type can contain sub-chunks, the \texttt{Read} method is called recursively using the new object as the mother.  
The number of bytes to read is computed from the \texttt{chunk size} and from the number of byte already read.
The example below shows what is done when a chunk of Id \texttt{C\_FEB}\footnote{\texttt{C\_FEB} is the \textit{Id} associated to the \texttt{FROG\_Element\_Base}.} is found in the file:

\begin{small}
\begin{verbatim}
case C_FEB:
   TemporyElement = new FROG_Element_Base(chunk_type);                          
   mother->addDaughter(TemporyElement);
   chunk_read += Read(pFile,TemporyElement,chunk_size-chunk_read);      
   break;
}
\end{verbatim}
\end{small}

When the chunks contains \textit{data} in addition of sub-chunks, the data are read using the object \texttt{constructor} with the file pointer in argument, the number of bytes read by the constructor has to be counted as well.
The \texttt{FROG\_Element\_Base\_With\_DetId} is a perfect example of such an object, the code below describes what is done when a chunk \textit{Id} of that type is found : 
\begin{small}
\begin{verbatim}
case C_FEB_DETID:
   TemporyElement = new FROG_Element_Base_With_DetId(chunk_type,pFile);                         
   mother->addDaughter(TemporyElement);
   chunk_read += FROG_Element_Base_With_DetId::sizeOf()-6;
   chunk_read += Read(pFile,TemporyElement,chunk_size-chunk_read);
   break;
\end{verbatim}
\end{small}

The code below is used when an unknown chunk \textit{Id} is found, which generally indicates a file corruption or incompatibilities.  
The unknown chunk is simply skipped and a warning message is printed out.  
The program will anyway display the other objects.

\begin{small}
\begin{verbatim}
default:
   printf("Unknown ChunkID (%i)\n",chunk_type);
   printf("Chunk will be skipped\n");
   printf("The program may be not running properly\n");
   fseek (pFile,chunk_size-chunk_read,SEEK_CUR);
   chunk_read += chunk_size-chunk_read;
   break;
\end{verbatim}
\end{small}

\paragraph*{FROG\_Element\_Primitive\_Line}
~~\\

This section details as an example the implementation of a leaf: \texttt{FROG\_Element\_Primitive\_Line} object that is used to draw a line segment in a 3D space.  This object is a leaf since it contains \textit{data} but no sub-chunks.
It can be used for the detector geometry design but also for the events representation, for example, a track in a detector without magnetic field is generally displayed as a straight line.
The class and variables declaration are done with this code: 

\begin{small}
\begin{verbatim}
#include "FROG_Element_Base_With_DetId.h"

class FROG_Element_Primitive_Line:
public FROG_Element_Base_With_DetId {
 public:	
    float P1X;    float P1Y;  float P1Z;
    float P2X;    float P2Y;  float P2Z;	
    FROG_Objects* frogObjects_;
\end{verbatim}
\end{small}

\texttt{float} variables are the positions of the two points of the line in cartesian XYZ coordinates.
The \texttt{FROG\_Objects*} variable is a container for few global parameters.
Since \texttt{FROG\_Element\_Primitive\_Line} inherits from \texttt{FROG\_Element\_Base\_With\_DetId}, the line object has also a \texttt{detId} that can be used to access the object to define its style for instance. 
The method \texttt{isCompactible} returns \texttt{true} if the object has a fixed \textit{size} and the method \texttt{sizeOf} returns the \textit{size} of an unique chunk.
This object has a fixed \textit{size} of $34~\textrm{Bytes}$\footnote{SizeOf(EmptyChunk) + 6*SizeOf(float) + SizeOf(DetId) = 6}.

\begin{small}
\begin{verbatim}
    virtual bool isCompactible(){
       return true; 
    }
    
    static unsigned int sizeOf(){
       return FROG_Element_Base_With_DetId::sizeOf() + 6*sizeof(float); 
    }
\end{verbatim}
\end{small}
The class \texttt{constructor} is defined below:
\begin{small}
\begin{verbatim}
    FROG_Element_Primitive_Line(
    unsigned int detId,			  
    float p1X,   float p1Y,   float p1Z,
    float p2X,	 float p2Y,   float p2Z):
    FROG_Element_Base_With_DetId(C_PRIMITIVE_LINE,detId),
    P1X(p1X),    P1Y(p1Y),    P1Z(p1Z),
    P2X(p2X),    P2Y(p2Y),    P2Z(p2Z)	
    {    
       size_ = sizeOf();
    }
\end{verbatim}
\end{small}

The arguments of the \texttt{FROG\_Element\_Base\_With\_DetId} constructor are the chunk type (\texttt{C\_PRIMITIVE\_LINE}) and the \texttt{detId}.  
The class needs a second constructor that build the object from the input file. 
\newpage

\begin{small}
\begin{verbatim}
    FROG_Element_Primitive_Line(FILE* pFile):
    FROG_Element_Base_With_DetId(C_PRIMITIVE_LINE)
    {   
       size_ = sizeOf();
       fread(&detId_           ,sizeof(detId_)         ,1,pFile);				
       fread(&P1X              ,sizeof(P1X)            ,1,pFile);
       fread(&P1Y              ,sizeof(P1Y)            ,1,pFile);
       fread(&P1Z              ,sizeof(P1Z)            ,1,pFile);
       fread(&P2X              ,sizeof(P2X)            ,1,pFile);
       fread(&P2Y              ,sizeof(P2Y)            ,1,pFile);
       fread(&P2Z              ,sizeof(P2Z)            ,1,pFile);
    }
\end{verbatim}
\end{small}

Data have to be read/written in the same order.  
Since the object do not contain daughters, the \texttt{write()} is easyer than \texttt{FROG\_Element\_Base\_With\_DetId::write()}.
\begin{small}
\begin{verbatim}
    virtual void write () {
       size_ = sizeOf();
       data_ = new unsigned char[size_-6];
       data_ = FillBuffer( data_, &detId_,   sizeof(detId_));
       data_ = FillBuffer( data_, &P1X,      sizeof(P1X));
       data_ = FillBuffer( data_, &P1Y,	     sizeof(P1Y));
       data_ = FillBuffer( data_, &P1Z,	     sizeof(P1Z));
       data_ = FillBuffer( data_, &P2X,	     sizeof(P2X));
       data_ = FillBuffer( data_, &P2Y,	     sizeof(P2Y));
       data_ = FillBuffer( data_, &P2Z,	     sizeof(P2Z));
       data_ = (unsigned char*)((unsigned long)data_ - (size_-6) );		
    }
\end{verbatim}	
\end{small}

The \texttt{data\_} array is directly initialised by the \texttt{write} function.
The object data are iteratively copied in the \texttt{data\_} variable with the \texttt{FillBuffer} function. 
The buffer cursor is moved back to the beginning of the buffer.

The \texttt{init} is generally called two times.  At the first time, the argument of the function will points to a \texttt{FROG\_Objects*}, while at the second call the pointer is \texttt{NULL}.

\begin{small}
\begin{verbatim}
    virtual void init(void* frogObjects){
       if(frogObjects!=NULL){
          frogObjects_ = (FROG_Objects*) frogObjects;			
       }else if(frogObjects==NULL && frogObjects_ != NULL){
          if(mother_!=NULL&&mother_->style_ !=NULL){
             style_ = new FROG_Objects_Style(mother_->style_);
          }else{
             style_ = new FROG_Objects_Style();	
          }	

          frogObjects_->frogCard_->GetColor(
             style_->color_     ,"Id_%i_Color"	  	,detId_);
          frogObjects_->frogCard_->GetFloat(
             &style_->thickness_ ,"Id_%i_Thickness"	,detId_);
       }
    }
\end{verbatim}	
\end{small}

If the method's argument is \texttt{NULL}, the \texttt{init} reads the colour and thickness parameters from the configuration file, using \texttt{frogObjects\_->frogCard\_->Get...} and store them into the \texttt{style\_}.
If no parameters are given, the default colour and thickness value of the mother is used. 
Finally the \texttt{display} method is almost only \textsc{OpenGL} code:
\newpage

\begin{small}
\begin{verbatim}
#ifdef FROG_OPENGL
    virtual void display(bool UseDisplayList=true, float* color=NULL){
       // first Init Colors & Style!
       init(NULL);

       glLineWidth(style_->thickness_);
       if(color){glColor4fv(color);}else{glColor4fv(style_->color_);}
       glLoadName( detId_ );

       glBegin (GL_LINES); 			
          glVertex3d (P1X, P1Y, P1Z);
          glVertex3d (P2X, P2Y, P2Z);
       glEnd(); 		
    }
#endif
\end{verbatim}	
\end{small}

First, the style is set by \texttt{init(NULL)}.
The right colour and thickness to used are transmitted to \textsc{OpenGL} for the line drawing, \texttt{glLoadName} is used to give an \textsc{OpenGL} name to the displayed line, this is need to allow mouse selection of the object (see sec.~\ref{sec:mouseInterface}).
Finally, the line is drawn by giving the two vertex positions between the markup \texttt{glBegin(GL\_LINES)} and \texttt{glEnd()}.

An extra method can be added to display some text when the object is mouse selected.
\begin{small}
\begin{verbatim}
    virtual void printInfos(char* buffer){		
       sprintf(buffer,"LINE : Hit1 : x=%f y=%f z=%f",P1X,P1Y,PIZ);
       sprintf(buffer,"LINE : Hit2 : x=%f y=%f z=%f",P2X,P2Y,P2Z);
    }
\end{verbatim}
\end{small}

The function just fill a text buffer with the text to print on screen when the object is selected.

These are all the functions an advanced user needs to implement in order to add its own objects toolkit.
More than 25 object classes can be found in \texttt{FROG/FROG\_Element\_*.h} that can be used as object class examples.

\subsubsection*{Geometry and Events Classes}

These two classes handle the \textsc{Frog} Elements tree, which is one of the private data members.
The classes also contains specific functions useful to display only few parts of the geometry and/or the events.  
In order to keep the random-access memory used by the Displayer reduced, the event class loads the events one by one.  
When the next event has to be drawn, the current event is firstly unloaded and only then, the next one is loaded.
This technique is not applicable for the geometry data because the event processing can request for a detector part information at any time.  So it is more efficient to load the entire detector geometry at the beginning and to keep it in memory up to the end.

These two classes will also take the \textsc{Frog} element tree writing in charge.  
The basic \textsc{Frog} user is not supposed to deal with the \texttt{FROG\_Element\_Base}.  
Those are just internal format used by the Geometry and Events classes to read, write and display objects.  
This fact will be shown in the next section.

\newpage

\subsection*{Appendix B: Complete Example}
\addcontentsline{toc}{subsection}{Appendix B: Complete Example}

%\section{Complete Example}
\label{sec:Complete_Example}

This section shows an example of code used for a fictitious tracking experiment composed of a particle source beam, eleven tracking layers and a block to stop the beam.
Others examples/tutorials are also available online  as a tutorial~\cite{REF_FROGTUT}.

\subsubsection*{Producer Code}

The Producer code is divided into four different methods: \texttt{main}, \texttt{Build\_Geometry} creates and stores the geometry of the detector, \texttt{Build\_Events} and \texttt{Build\_1Event} simulate and store the events.
The code includes the needed \textsc{Frog} classes, and defines the functions implemented in the file. 
Eventually, the main is implemented.

\begin{small}
\begin{verbatim}
#include <stdio.h>
#include <iostream>
#include <math.h>

#include "FROG/FROG_DetId.h"
#include "FROG/FROG_Geometry.h"
#include "FROG/FROG_Events.h"
#include "FROG/FROG_Element_Tools.h"

void Build_Geometry();
void Build_Events();
void Build_1Event(FROG_Element_Event* event);

int main()
{   
  Build_Geometry();   
  Build_Events();
}
\end{verbatim}
\end{small}

The \texttt{Build\_Geometry function} creates the tree structure that is stored in the file.
The root of the tree is necessary a \texttt{FROG\_Element\_Base} with a \texttt{C\_PRIMARY} chunk Id.
The geometry branch is created and attached to this primary object.
Other branches are created to reflect the detector structure, since they are of type \texttt{FROG\_Element\_Base\_Width\_DetId\_And\_Name}, those branches contain a unique detId that will be used to access the branch, and a string that give a name to the branch (e.g., "TRACKER" or "OTHER").
The eleven tracking layers are created in a loop and are represented by a rectangle attached to the "TRACKER" branch.

\begin{small}
\begin{verbatim}
void Build_Geometry()
{	
  FROG_Element_Base* prim   = new FROG_Element_Base(C_PRIMARY);
  FROG_Element_Base* mygeom = new FROG_Element_Base(C_GEOMETRY); 	
  prim->addDaughter(mygeom);

  FROG_Element_Base_With_DetId_And_Name* tracker;
  tracker = new FROG_Element_Base_With_DetId_And_Name(9000000,"TRACKER");	
  mygeom->addDaughter(tracker);

  unsigned int DetIdCount = 1;

  // TRACKING LAYERS
  for(int i=0;i<=10;i++)
  {		
    FROG_Element_Primitive_Rectangle* layer;
    layer = new FROG_Element_Primitive_Rectangle(
                           400000000+DetIdCount,  // DetId
                           0  ,0  ,50*i,          // Position
                           50 ,0  ,0   ,          // Width
                           0  ,50 ,0   );	  // Length	
    tracker->addDaughter(layer);
    DetIdCount++;
  }
\end{verbatim}  
\end{small}

The ending block that stops the particles and the particle source are attached to the "OTHER" branch.  They are both represented by a cuboid volume.
Finally, a \textsc{Frog\_Geometry} object is created from the root of the data structure.
The \textsc{Frog\_Geometry::save()} creates and fill the ouput file of name \texttt{"MyCustomTracker.geom"}.

\begin{small}
\begin{verbatim}  
  FROG_Element_Base_With_DetId_And_Name* others;
  others = new FROG_Element_Base_With_DetId_And_Name(8000000,"OTHER");	
  mygeom->addDaughter(others);

  // END BLOCK
  FROG_Element_Primitive_Cube* End = new FROG_Element_Primitive_Cube(
                           400000000+DetIdCount,  // DetId																		  
                           0  ,0  ,600 ,          // Position
                           50 ,0  ,0   ,          // Width
                           0  ,50 ,0   ,          // Length
                           0  ,0  ,50  );         // Thickness
  others->addDaughter(End);   DetIdCount++;


  // PARTICLE GUN
  FROG_Element_Primitive_Cube* PG = new FROG_Element_Primitive_Cube(
                           400000000+DetIdCount,  // DetId																			  
                           0  ,0  ,-300,          // Position
                           5  ,0  ,0   ,          // Width
                           0  ,5  ,0   ,          // Length
                           0  ,0  ,50  );         // Thickness
  others->addDaughter(PG);   DetIdCount++;

  FROG_Geometry* CustomGeom = new FROG_Geometry(prim);
  CustomGeom->Save("MyCustomTracker.geom");
  delete CustomGeom;

  return;
}
\end{verbatim}
\end{small}

The \texttt{Build\_Events} method creates the data tree that will contain all the new events.
Hundred events are created and push in this tree.
The events are of type \texttt{FROG\_Element\_Event}, the run and event numbers are given as an argument of the constructor.
Once the new event has been filled by \texttt{Build\_1Event}, and pushed in the data tree, they are stored on disk (one by one) by \texttt{FROG\_Events::SaveInLive}.
The arguments of the function are used to set : the output file name, if the file has to be gzipped, if it has to be closed, and the maximum file size in bytes.
The return value is \texttt{FALSE} when the new event will make the file size exceeding the maximum allowed file size ($2~\textrm{MB}$ in this case).
At the end, \texttt{SaveInLive} is call a last time to force the closing of the file.

\begin{small}
\begin{verbatim} 
void Build_Events()
{
  FROG_Events* events = new FROG_Events();
  for(unsigned int i=0;i<100;i++){
    FROG_Element_Event* event =  new FROG_Element_Event(1,i);            
    Build_1Event(event);            
    events->AddEvent(event);			
    if(!events->SaveInLive("SimulatedEvents.vis",false,false,2000000))
    {
      printf("File Size Exeeded 2.000.000 bytes\n");
      printf("Stop here\n");
      exit(0);
    }			
  }
  events->SaveInLive("SimulatedEvents.vis",true,false,2000000);
  delete events;
}
\end{verbatim}
\end{small}

\newpage

The method \texttt{Build\_1Event} simulates and stores one event.
The simulation consists a straight line propagation of the particle using a random emission angle from the particle source.
During the propagation, a new hit is created and attached to the data tree, each time the particle is crossing a tracking layer.
The data structure is such that the hits are contained in a track itself contained in a track collection.

\begin{small}
\begin{verbatim}
void Build_1Event(FROG_Element_Event* event)
{
  // Particle Gun direction and strength
  double phi   = (rand()%360)*0.01745;
  double theta = (rand()%8)  *0.01745;
  double P     = 10;

  // Particle Gun position
  double pos_x = 0;
  double pos_y = 0;
  double pos_z = -300;

  // Particles direction (cartesian coord)
  double dir_x = P*cos(phi)*sin(theta);
  double dir_y = P*sin(phi)*sin(theta);
  double dir_z = P*cos(theta);
  double Pt    = sqrt(dir_x*dir_x + dir_y*dir_y);

  FROG_Element_Base_With_DetId_And_Name* Tracks;
  Tracks = new FROG_Element_Base_With_DetId_And_Name(EVTID_TRK, "Track Collection");	
  FROG_Element_Event_Track* track= new FROG_Element_Event_Track(0,P,Pt,0);	

  // Propagation loop
  double dt = 0.1;
  for(unsigned int i=0;i<10000;i++)
  {
    pos_x += dir_x * dt;
    pos_y += dir_y * dt;
    pos_z += dir_z * dt;

    if(pos_z<0 || pos_z>=600) continue;

    // Is the particle crossing a tracking layer --> if yes, creates a hit
    if(((int)pos_z)%50==0 && fabs(pos_x)<=50 && fabs(pos_y)<=50)
    { 			
      FROG_Element_Event_Hit* hit = new FROG_Element_Event_Hit(
                                    400000001 + ((int)pos_z)/50,// DetId
                                    pos_x, pos_y, pos_z,        // position
                                    -1);                        // dEdx
      track->addDaughter(hit);						
    }
  }

  Tracks->addDaughter(track);
  event->addDaughter(Tracks);
  
  return;	
}
\end{verbatim}
\end{small}

This simple example shows how a user can display his experimental setup using \textsc{Frog} in only few lines of code.
The standard \textsc{Frog} package contains already more than 25 objects that can be used to render the detector geometry and events.  Those classes have been shown in the Figure~\ref{fig:CORE_ClassDiagram1}.
In addition of those several available classes, the users can also create new classes to display particular objects that fit his needs, a tutorial is available online~\cite{REF_FROGTUT}.

\newpage

\subsubsection*{Configuration Code}

The ASCII code below is the configuration file used to run \textsc{Frog} on the fictitious tracking experiment.

\begin{small}
\begin{verbatim}
InputVisFile          = {SimulatedEvents.vis};      // Input event file
InputGeom             = {MyCustomTracker.geom};     // input geometry file

GeomToDisplay         = {9000000 , 8000000};
EventToDisplay        = {23100000};                 // Tracks

Event_Number          = 0;                          // first event number
Event_Time            = 30;                         // event will change every 30 Sec
Geometry_WireFrame    = true;			    // Geometry is draw as wireframe
Screenshot_Format     = png;                        // Screenshot format

BackGround_Color      = {1.0 , 1.0 , 1.0 };         // White
Txt_       Color      = {0.0 , 0.0 , 0.0 , 1.0};    // Black
ZAxis     _Color      = {1.0 , 0.5 , 1.0 , 0.3};    // Red+Blue
ZAxis _Thickness      = 0.0;

Id_23100000_Color     = { 1.0 , 0.0 , 0.0 , 1.0 };  // Tracks
Id_23100000_Thickness = 4.0;                        // Tracks
Id_23100000_Marker    = 100;                        // Tracks
Id_23100000_MarkerSize= 5;                          // Tracks
Id_9000000_Color      = { 0.4 , 0.4 , 1.0 , 0.5 };  // Tracking Layers
Id_8000000_Color      = { 0.4 , 0.4 , 0.4 , 1.0 };  // End Block

ActiveViews           = {View3D, View2DZ, View2DR, ViewMenu};

View3D_Type           = 3D;
View3D_Viewport_X     = 0.0;
View3D_Viewport_Y     = 0.3;
View3D_Viewport_W     = 1.0;
View3D_Viewport_H     = 0.8;
View3D_Cam_Pos_Theta  = 0.3;
View3D_Cam_Pos_Phi    = 0;
View3D_Cam_Pos_R      = 500;

View2DZ_Type          = 2D;
View2DZ_Viewport_X    = 0.4;
View2DZ_Viewport_Y    = 0.0;
View2DZ_Viewport_W    = 0.6;
View2DZ_Viewport_H    = 0.2;
View2DZ_Cam_Pos_Phi   = 0.01;
View2DZ_Cam_Pos_R     = 600;
View2DZ_Slice_Depth   = 800;

View2DR_Type          = 2D;
View2DR_Viewport_X    = 0.1;
View2DR_Viewport_Y    = 0.0;
View2DR_Viewport_W    = 0.2;
View2DR_Viewport_H    = 0.2;
View2DR_Cam_Pos_Phi   = 1.57;
View2DR_Cam_Pos_R     = 100;
View2DR_Slice_Depth   = 1000;

ViewMenu_Type         = Menu;
ViewMenu_Viewport_X   = 0.15;
ViewMenu_Viewport_Y   = 0.05;
ViewMenu_Viewport_W   = 0.7;
ViewMenu_Viewport_H   = 0.9;
ViewMenu_Alpha        = 0.8;
\end{verbatim}
\end{small}
\newpage

\bibliographystyle{ieeetr} 
\bibliography{Biblio}

\begin{thebibliography}{10}

\bibitem{REF_FROG}
L.~{Quertenmont} and {Roberfroid}, ``{\bf \textsc{Frog}: the Fast and Realistic
  OpenGL Displayer},'' \newline\url{http://projects.hepforge.org/frog/}.

\bibitem{REF_OPENGL}
``{\bf \textsc{OpenGL}: Open Graphical Library},''
  \newline\url{http://www.opengl.org}.

\bibitem{REF_GLUT}
``{\bf \textsc{Glut}: \textsc{OpenGL} Utility Toolkit},''
  \newline\url{http://www.opengl.org/resources/libraries/glut/}.

\bibitem{REF_CMS}
M.~Della~Negra, A.~Petrilli, A.~Hervé, and al., ``{\bf CMS: The Compact Muon
  Solenoid}, cms physics: Technical design report 1,'' 2006.
\newblock {\em CERN/LHCC 2006-001}.

\bibitem{REF_FROGTUT}
L.~{Quertenmont} and {Roberfroid}, ``{\bf \textsc{Frog} Tutorials},''
  \newline\url{http://projects.hepforge.org/frog/Tutorial/Tutorial.html}.

\bibitem{REF_LIBCURL}
``{\bf LIBCURL},'' \newline\url{http://curl.haxx.se/}.

\bibitem{REF_ZLIB}
``{\bf ZLIB},'' \newline\url{http://www.zlib.net/}.

\bibitem{REF_GL2PS}
``{\bf \textsc{GL2PS}: an \textsc{OpenGL} to PostScript printing library},''
  \newline\url{http://www.geuz.org/gl2ps/}.

\bibitem{REF_PNG}
``{\bf PNGLIB: Portable Network Graphics Library},''
  \newline\url{http://www.libpng.org/}.

\bibitem{REF_GASTOF}
L.~Bonnet, T.~Pierzchala, K.~Piotrzowski, and P.~Rodeghiero, ``{\bf
  \textsc{GasTOF}: Ultra-fast ToF forward detector for exclusive processes at
  the LHC},'' 2006.
\newblock {\em arXiv:hep-ph/0703320; CP3-06-18}.

\bibitem{REF_DELPHES}
S.~Ovyn and X.~Rouby, ``{\bf \textsc{Delphes}}.''

\bibitem{REF_GASTOF2}
L.~Bonnet and al., ``{\bf \textsc{GasTOF}: Paper in preparation},'' 2009.

\bibitem{REF_ATLAS}
T.~A. Collaboration, ``{\bf ATLAS} detector and physics: Performance technical
  design report,'' 1999.
\newblock {\em CERN/LHCC 1999-14/15}.

\bibitem{REF_FP420}
M.~Albrow, R.~Appleby, and al., ``{\bf FP420: The FP420 R\&D Project: Higgs and
  New Physics with forward protons at the LHC},'' 2008.
\newblock {\em arXiv:hep-ex/08060302}.

\bibitem{REF_LHC}
M.~Benedikt, P.~Collier, V.~Mertens, J.~Poole, and K.~Schindl, {\em LHC Design
  Report}.
\newblock Geneva: CERN, 2004.

\bibitem{REF_GASTOF_SIM}
T.~Pierzchala and N.~Schul, ``{\bf \textsc{GasTOF}} simulator.''

\bibitem{REF_CMSSW}
M.~Della~Negra, A.~Petrilli, A.~Hervé, and al., ``{\bf CMS: The Compact Muon
  Solenoid}, cms physics: Technical design report 2,'' 2006.
\newblock {\em CERN/LHCC 2006-021}.

\bibitem{REF_CALTECH}
``{\bf Caltech Grid Enabled Analysis},''
  \newline\url{http://ultralight.caltech.edu/web-site/gae/html/index.html }.

\bibitem{REF_DLIST}
``{\bf \textsc{OpenGL} Display List},''
  \newline\url{http://www.glprogramming.com/red/chapter07.html}.

\end{thebibliography}

\end{document}